# HOPE: An Automatically Differentiable High-Order Non-Oscillatory Finite-Volume Shallow-Water Dynamic Core

Zhou Lilong[1,2]

1. *Tsinghua University, Beijing, China*
2. *CMA Earth System Modeling and Prediction Centre (CEMC), Beijing, China*

**Corresponding author:** Zhou Lilong (zhoull@cma.gov.cn)

# Abstract

An automatically differentiable, high-order non-oscillatory finite volume shallow water dynamic core has been constructed on a cubed sphere grid. This dynamic core has four advantageous properties: high order accuracy, essential non-oscillation, mass conservation, and scalability. Besides, the code development is based on PyTorch, enabling the model to run seamlessly on both CPU and GPU, while naturally possessing the capability of automatic differentiation. We named the new dynamic core as High Order Prediction Environment (HOPE). The spatial reconstruction is based on the two-dimensional tensor product polynomial (TPP) and the genuine two-dimensional Weighted Essentially Non-Oscillatory (WENO) scheme. A novel panel boundary approach ensures that the accuracy can reach arbitrary order. These algorithms have very high degree of compatibility with GPU architecture, allowing the computational overhead to be mitigated through the utilization of GPU. The LMARS (Low Mach number Approximate Riemann Solver) scheme is adopted as Riemann solvers to determine fluxes on the Gaussian points on edges. Flux across the interface between each cell edge is computed using Gaussian quadrature, and the tendencies of prognostic variables are obtained by integration all the source terms and the fluxes across the cell boundaries. This shallow water dynamic core exhibits outstanding performance in ideal shallow water test cases. In the steady-state geostrophic flow, the $11^{th}$ order scheme reduces errors to nearly double precision round-off error even on coarse grids (1°×1°). Furthermore, HOPE maintains the Rossby-Haurwitz wave over 100 days without collapse. To test the non-oscillation property, we designed a cylinder dam break case, the WENO approach effectively suppresses non-physical oscillation, and the genuine two-dimensional reconstruction exhibits significantly better isotropy than the dimension-by-dimension scheme.

# 1. Introduction

Numerical weather prediction (NWP) became a foundation of weather forecasting in past decades. The horizontal resolution of operational models in each NWP center is now finer than 10km, European Centre for Medium-Range Weather

Forecasts (ECMWF) enabled real-time global weather forecasting with about 9km horizontal grid since 2016[24], and Numerical Weather Prediction Center (NWPC) of China Meteorological Administration (CMA) increased the operational regional model resolution to 3km in 2017[10]. Finer resolution brings us not only more details about atmosphere, but also more challenge about simulating small synoptic systems, steeper topography and especially more computational cost.

In recent years, the machine learning (ML) technology has been wildly utilized in atmospheric numerical simulations. Some research suggests that ML can overcome the challenge of computational cost and even improve the forecast accuracy. There are several common methods of leveraging ML in weather prediction. A simple and effective method is to apply machine learning to post-processing of model data, using neural networks to correct the model forecast fields so that the forecast results are closer to actual observations[42]. However, this type of method has limited effectiveness in correcting forecast results. Since it uses model forecasts as the source data, when the model forecasts have significant deviations, the correction effectiveness also diminishes.

Another method is establishing a data driven NN model to surrogate the entire prediction process, such as Pangu-Weather[3], FengWu[6], GraphCast[16], NowcastNet [45] and so on. The NN models perform excellent forecasting accuracy for the large-scale atmosphere state, meanwhile they are thousand times faster than traditional numerical mode. However, training these models require amount of reanalysis data, and the training process is very expensive, it costs hundreds of GPU to execute the training over weeks, and the forecasting becomes blurrier along with the leading time increase.

Some researchers attempted to merge the traditional NWP model and NN into a hybrid model. In the traditional NWP model, the solving process of governing equation is separated to dynamic part and physical part. The purpose of dynamic part is solving the grid-scale dynamic partial differential equation (PDE) by numerical methods, i.e. finite volume (FV), finite difference (FD), discontinuous Galerkin (DG) and so on. The physical part deals with the sub-grid physical processes by parameterization, which causes significant larger uncertainties and errors than dynamic part [30]. However, machine learning algorithms happen to be well-suited for addressing such problem. Wang et al. (2022) [39] emulating physical parameterization through a surrogate model, then coupling with the numerical dynamic core. The surrogate model is trained offline, requiring the long-term execution of the original numerical model to extract input and output data from the physical parameterization module for use as labels. The offline training of physical parameterization module needs much less data than the full model surrogation scheme. However, during the time marching, the prediction error emerges as a nonlinear superposition of the dynamic error, which arises from solving PDE, and the physical error, stemming from the distortion introduced by parameterizing the physical process. Offline training is a purely data-driven approach, where the surrogate model lacks awareness of the underlying dynamic core's behavior.

A more thorough solution would be to develop the entire numerical model on a machine learning programming platform, such as TensorFlow or PyTorch[15]. In this scenario, the dynamic core is based on a traditional numerical PDE solver, while the physical parameterization module is a neural network (NN) trained using the backpropagation of prediction residuals. Unlike the second method, this approach couples the dynamic core error and the physical parameterization error through backpropagation. Therefore, during the training process, the NN-based physical parameterization module can obtain more comprehensive residual information. NeuralGCM [15] proposed a hybrid model by combining a spectral numerical dynamic core and NN based physical parameterization model. The dynamic core based on governing equations imposes rigorous physical constraints on the model, which eliminates the blurriness present in purely data-driven models within the NeuralGCM framework. Additionally, the power spectra performance of NeuralGCM is superior to that of purely data-driven meteorological models.

For NeuralGCM, although the spectral dynamic core can provide theoretical infinite accuracy, the inherent shortcomings of the spectral model still persist. Specifically, it fails to preserve mass conservation, and the global nature of spectral expansion also restricts the scalability of this method.

To overcome above problems, we introduce a shallow water dynamic core named High Order Prediction Environment (HOPE). The contributions of this study are

1) We develop HOPE, a shallow water model has four advantageous properties: arbitrary high order accuracy, essential non-oscillation, mass conservation, and scalability.

2) We desire a novel high order ghost cell interpolation scheme for cubed sphere grid, it needs only one sparse matrix multiplication to reach arbitrary odd convergence order.

3) We implement genuine two-dimensional reconstruction on cubed sphere, comparing to the dimension-by-dimension scheme, the genuine two-dimensional provides less dimension split error.

4) HOPE is developed on PyTorch, the auto-differentiate capability is naturally obtained, it's easy to couple with any neural-network (NN) based functions, such as sub-grid physical parameterization.

5) The algorithms of the HOPE model primarily involve convolution and matrix multiplications, which are widely used in the artificial neural networks and highly compatible with GPUs. HOPE demonstrates excellent computational efficiency on GPU platforms.

In the following part of the introduction, we introduce the relevant work on constructing the HOPE model, and from this, we elaborate on the challenges and motivations for establishing the algorithm of the dynamic core. High-order accuracy is an extremely appealing trait for the design of a dynamic core, particularly in high-resolution atmospheric simulations. A dynamic core model with high-order accuracy produces significantly less simulation error in smooth regions compared to a low-order model. Furthermore, even when the resolution is equivalent or coarser, a high-order model is capable of resolving finer details than a low-order one. There are plenty of researches about implementing high order schemes in spherical shallow water model, Chen and Xiao[1] introduced a multi-moment finite volume (MCV) scheme with 3rd and 4th order accuracy to develop the shallow water model on a cubed sphere. Ii and Xiao[11] extended MCV based shallow water model to icosahedral grid with $3^{rd}$ and $4^{th}$ order accuracy. Katta et al.[13][14] compared 1D and 2D reconstructions on a cubed sphere, they found that the 1D scheme reduced the accuracy to 2nd order even when using a 5th order dimension-by-dimension Weighted Essentially Non-Oscillatory (WENO) reconstruction. In contrast, the fully 2D scheme maintained high-order accuracy in smooth shallow water test cases, albeit with a higher computational cost compared to the 1D scheme. Both Ullrich et al. (2010) [37] and our tests confirm this conclusion, indicating that a dimension-by-dimension scheme is not a reliable choice for establishing a high-order finite volume dynamic core on cubed sphere grid. Furthermore, the absence of cross-derivative terms results in unrealistic anisotropy, as demonstrated in the cylinder dam break case discussed in section 5.

Ullrich et al.[37][38] developed a high-order finite volume model on cubed sphere, MCORE. The horizontal reconstruction process unfolds in two stages: initially, cell center values are determined through a convolution operator, followed by the application of a finite difference operator to compute derivatives. MOCRE attains $(k-1)^{th}$ order accuracy when employing a stencil width of $k$. The authors assert that MCORE's convergence accuracy can theoretically be of arbitrary order. However, in our practical numerical tests, we found that the accuracy does not surpass the 7th order. This limitation arises when using a one-sided ghost interpolation scheme, which leads to numerical oscillations originating from the corner zones of the panels when the stencil size is 9×9 or larger.

In this article, we devise the reconstruction based on tensor product polynomial (TPP). When the stencil width is $k$, our method achieves $k^{th}$ order accuracy, surpassing MCORE by one order of accuracy with the same stencil width. In addition, we have developed a new class of ghost interpolation schemes that abandon the use of one-sided stencils and instead adopt central stencils. This new approach enables the scheme to overcome the non-physical oscillations arising from interpolation at panel boundaries. Our method allows for arbitrary accuracy, and we have verified this by testing up to the 13th order.

From the properties of the Taylor series, we note that it can only effectively approximate a function when the higher-order derivatives exist and the series converges. When the continuity of the field is poor, the higher-order derivatives do not exist, or the series residuals increase with the order, using a higher-order series will actually introduce greater errors. Therefore, for reconstruction schemes based on polynomial functions, high-accuracy schemes should only be adopted when the field is sufficiently smooth. For fields with poor continuity, the accuracy should be appropriately reduced to ensure the effectiveness of the reconstruction.

WENO is an adaptive numerical scheme widely used in computational fluid dynamics (CFD) simulations. It effectively eliminates non-physical oscillations caused by sharp discontinuities without compromising accuracy in smooth regions [12][21]. Previous research on implementing WENO in atmospheric simulations has demonstrated its attractive advantages. For instance, in the density current test case [32], high-order WENO schemes achieved convergent solutions even at coarse resolutions, while centered schemes failed to do so [31]. Lunet et al. (2017) [22] combined WENO with Explicit Runge-Kutta methods in the Meso-NH model, resulting in more stable and non-oscillatory transitions with sharp discontinuities compared to centered schemes. Furthermore, when compared to the fourth-order centered scheme with leapfrog time marching, the new algorithm that combines WENO with explicit Runge-Kutta method (ERK) improved computational efficiency by over three times. In Norman's colliding thermals test case [25], non-physical oscillations were observed without the use of a WENO limiter, whereas the WENO scheme produced a more reasonable result, even when the perturbation gradient was steep.

The original WENO was developed for the one-dimensional case [21]. Subsequently, Shi et al.[33] extended it to two dimensions using two different methods: a genuine 2D (WENO2D) scheme and a dimension-by-dimension scheme. The genuine 2D scheme yielded lower error but required more computational time. From another perspective, the genuine 2D scheme is more suitable for complex geometric discretization. Zhu and Shu (2018, 2019) [48][49][50] devised two-dimensional central WENO schemes for both regular and triangular meshes, where polynomial coefficients were determined by solving an overdetermined linear system using a least squares method. Zhao et al. further developed WENO for unstructured quadrilateral and triangular meshes[46].

Drawing inspiration from these advancements, we attempt to construct a dynamic core based on following considerations. Firstly, mass conservation is a fundamental property for a dynamic core. Secondly, the algorithm must be robust enough to handle both smooth and discontinuous fields. This requires the algorithm should maintain high-order accuracy in smooth regions and adapt to large gradients in non-smooth regions, the reconstruction polynomial need to be reduced to low order adaptively to eliminate non-physical oscillations. Lastly, given the rapid increase in computational cost with higher resolution, the dynamic core must be suitable for massively parallel computations.

We consider that combining the Finite Volume Method (FVM) with a genuine 2D reconstruction scheme and using a Riemann solver to calculate the flux can effectively meet all the aforementioned requirements. This is the core starting point of the HOPE algorithm. The FVM naturally conserves mass, and the genuine 2D reconstruction scheme based on TPP ensures high accuracy of the algorithm on cubed-sphere grids. When the WENO algorithm is employed, it achieves adaptability to non-smooth flow fields. Cell-centered numerical reconstruction may produce different results on either side of the cell boundary. In such cases, Riemann solvers can be used to determine the reconstruction information on both sides of the interface. Subsequently, the net flux can be obtained through Gaussian quadrature. All these schemes can be computed locally without the need for global communication, offering excellent scalability. Furthermore, in program implementation, the TPP-based reconstruction scheme can be directly represented as convolution operations, Gaussian quadrature can be expressed as matrix-vector multiplications, and the Riemann solver algorithm involves only basic arithmetic operations. The HOPE shallow-water dynamic core does not require any for-loops or if-branch judgments, making it highly suitable for GPU acceleration. Currently, we have implemented HOPE in both Fortran and PyTorch programming frameworks, with the PyTorch version offering clear advantages. Currently, we have implemented HOPE in both Fortran and PyTorch programming frameworks, with the PyTorch version exhibiting clear advantages. It leverages PyTorch's built-in high-performance functions for GPU acceleration and inherently possesses automatic differentiation capabilities. This makes it highly convenient for HOPE to couple with any neural network model in subsequent development without the need for additional development of backpropagation modules. An automatically differentiable dynamic framework also brings convenience to developing data assimilation framework. Traditionally, developing four-dimensional variational data assimilation systems required the prior development of tangent linear and adjoint models for the numerical model. However, as shown by the work of Xiao et al. (2023)[43], leveraging PyTorch's automatic differentiation capability eliminates the need for complex adjoint model development in four-dimensional variational data assimilation. This allows scientists to focus more on other meaningful research areas such as physical principles, high-precision algorithms, and dynamic-physical coupling methods.

The following content of this paper is presented as follows: in section 2, we introduce the cubed-sphere grid and the governing equations. Section 3 describes the numerical methods, including reconstruction schemes, panel boundary treatment method, and temporal marching scheme. In section 4, we introduce the method of high-performance implementation in PyTorch platform. In section 5, numerical experiments are conducted to demonstrate that the actual performance of the HOPE model aligns with its designed performance. Section 6 concludes this article and introduces our ongoing works and future planes. Section 7 is the appendix, we provide a detailed derivation of the new boundary treatment scheme.

## 2. Governing Equation on Cubed Sphere

Cubed sphere grid decomposes sphere to six panels, the computational space is structured and rectangular in each panel, these features make it easy to take high order reconstruction and massive threads parallel, details in Figure 2.1. The early research about solving primitive equation on cubed sphere can be found in Sadourny (1972)[32]. In recent decades, cubed sphere is used in different kinds of high order accuracy atmospheric models, Chen and Xiao[1] built a shallow water model by multi-moment constraint finite volume method on cubed sphere, $3^{rd}$ ~$4^{th}$ order accuracy was achieved. Ullrich et al.[37][38] developed a high order finite volume dynamic core based on cubed sphere, Nair et al.[26][27][28][29] established discontinuous Galerkin model on cubed sphere. In our research, cubed sphere is also adopted, even though the mesh is not orthogonal, we can still treat the computational space as rectangular grid by taking generalized curvilinear coordinate equation set. In this section, we introduce the shallow water equation set in generalized curvilinear coordinate, and special treatment of topography.

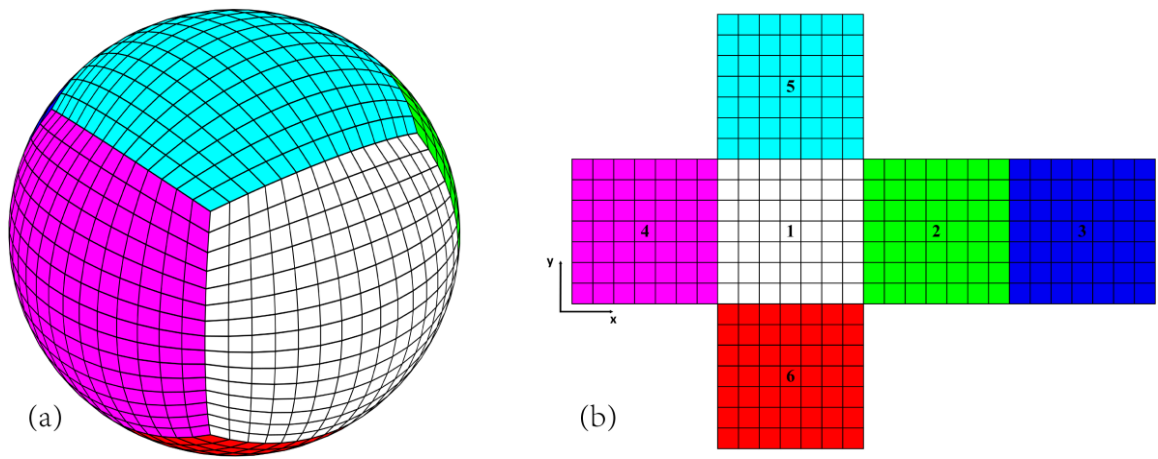

Figure 2.1 Cubed sphere grid. (a) Physical space; (b) Computational space. Six panels are identified by indices from 1 to 6.

Shallow water equation set on gnomonic equiangular cubed sphere grid is written as

$$\begin{cases} \dfrac{\partial \sqrt{G}\phi}{\partial t} + \dfrac{\partial \sqrt{G}\phi u}{\partial x} + \dfrac{\partial \sqrt{G}\phi v}{\partial y} = 0 \\ \dfrac{\partial \sqrt{G}\phi u}{\partial t} + \dfrac{\partial \sqrt{G}\left(\phi uu + \frac{1}{2}G^{11}\phi^2\right)}{\partial x} + \dfrac{\partial \sqrt{G}\left(\phi uv + \frac{1}{2}G^{12}\phi^2\right)}{\partial y} = \psi_M^1 + \psi_C^1 + \psi_B^1 \\ \dfrac{\partial \sqrt{G}\phi v}{\partial t} + \dfrac{\partial \sqrt{G}\left(\phi uv + \frac{1}{2}G^{21}\phi^2\right)}{\partial x} + \dfrac{\partial \sqrt{G}\left(\phi vv + \frac{1}{2}G^{22}\phi^2\right)}{\partial y} = \psi_M^2 + \psi_C^2 + \psi_B^2 \end{cases} \quad (2.1)$$

The gnomonic equiangular coordinates are represented by $(x, y, n_p)$, where $(x, y) \in \left[-\frac{\pi}{4}, \frac{\pi}{4}\right]$ are local equiangular coordinate of each panel and $n_p$ is panel number. $\phi = gh$ is geopotential height, $h$ is fluid thickness, $u, v$ is contravariant wind in $x, y$ direction, $g$ is gravity acceleration. $\psi_M, \psi_C, \psi_B$ are the metric term, Coriolis term and bottom topography influence term

$$\psi_M = \begin{pmatrix} \psi_M^1 \\ \psi_M^2 \end{pmatrix} = \frac{2\sqrt{G}}{\delta^2} \begin{pmatrix} -XY^2\phi uu + Y(1+Y^2)\phi uv \\ X(1+X^2)\phi uv - X^2Y\phi vv \end{pmatrix} \quad (2.2)$$

$$\psi_C = -\sqrt{G}\sqrt{G}f\mathbf{k} \times \phi\mathbf{u} = \sqrt{G}f \begin{pmatrix} -G^{12} & G^{11} \\ -G^{22} & G^{12} \end{pmatrix} \begin{pmatrix} \sqrt{G}\phi u \\ \sqrt{G}\phi v \end{pmatrix} \quad (2.3)$$

$$\psi_B = -\sqrt{G}\phi G^{ij}\frac{\partial \phi_s}{\partial x^j} = -\sqrt{G}\phi \begin{pmatrix} G^{11}\dfrac{\partial \phi_s}{\partial x} + G^{12}\dfrac{\partial \phi_s}{\partial y} \\ G^{21}\dfrac{\partial \phi_s}{\partial x} + G^{22}\dfrac{\partial \phi_s}{\partial y} \end{pmatrix} \quad (2.4)$$

where $X = \tan x$, $Y = \tan y$, $\delta = \sqrt{1 + X^2 + Y^2}$, and $f = 2\Omega\sin\theta$ is Coriolis parameter, and $\phi_s = gh_s$ is surface geopotential height, $h_s$ is surface height.

$$\sin\theta = \begin{cases} Y/\delta, & n_p \in \{1,2,3,4\} \\ 1/\delta, & n_p = 5 \\ -1/\delta, & n_p = 6 \end{cases} \quad (2.5)$$

The contravariant metric on cubed-sphere is

$$G^{ij} = \frac{\delta^2}{r^2(1+X^2)(1+X^2)} \begin{pmatrix} 1+Y^2 & XY \\ XY & 1+X^2 \end{pmatrix} \quad (2.6)$$

The covariant metric

$$G_{ij} = \frac{r^2(1+X^2)(1+Y^2)}{\delta^4} \begin{pmatrix} 1+X^2 & -XY \\ -XY & 1+Y^2 \end{pmatrix} \quad (2.7)$$

and the metric determinant is given by

$$\sqrt{G} = \sqrt{\det(G_{ij})} = \frac{r^2(1+X^2)(1+Y^2)}{\delta^3} \quad (2.8)$$

$r$ is radius of earth.

The contravariant wind vector $\mathbf{V} = (u, v)$ can be convert to wind vector on spherical LAT/LON coordinate $\mathbf{V}_s = (u_s, v_s)$ by the following formula

$$\begin{pmatrix} u_s \\ v_s \end{pmatrix} = J \begin{pmatrix} u \\ v \end{pmatrix} \quad (2.9)$$

where $J$ is a $2 \times 2$ conversion matrix, the expressions are different in each panel

$$J = r\begin{pmatrix} \cos\theta \frac{\partial \lambda}{\partial x} & \cos\theta \frac{\partial \lambda}{\partial y} \\ \frac{\partial \theta}{\partial x} & \frac{\partial \theta}{\partial y} \end{pmatrix} = \begin{cases} r\begin{pmatrix} \cos\theta & 0 \\ -\sin\theta\cos\theta\tan\lambda_p & \cos\lambda_p\cos^2\theta + \frac{\sin^2\theta}{\cos\lambda_p} \end{pmatrix}, & \text{panel } 1\sim 4 \\ r\begin{pmatrix} \cos\lambda\sin\theta\,\Gamma_1 & \sin\lambda\sin\theta\,\Gamma_2 \\ -\sin\lambda\sin^2\theta\,\Gamma_1 & \cos\lambda\sin^2\theta\,\Gamma_2 \end{pmatrix}, & \text{panel } 5 \\ r\begin{pmatrix} -\cos\lambda\sin\theta\,\Gamma_1 & \sin\lambda\sin\theta\,\Gamma_2 \\ \sin\lambda\sin^2\theta\,\Gamma_1 & \cos\lambda\sin^2\theta\,\Gamma_2 \end{pmatrix}, & \text{panel } 6 \end{cases} \quad (2.10)$$

$$\lambda_p = \lambda - \frac{\pi}{2}(i_{panel} - 1), \quad \Gamma_1 = 1 + \frac{\sin^2\lambda}{\tan^2\theta}, \quad \Gamma_2 = 1 + \frac{\cos^2\lambda}{\tan^2\theta} \quad (2.11)$$

where $\lambda, \theta$ are longitude and latitude, and $i_{panel}$ is the panel index as shown in Figure 2.1(b). The relation between $J$ and $G_{ij}$ is

$$G_{ij} = J^T J \quad (2.12)$$

In our numerical experiments, topography causes non-physical oscillation while we using equation set Eq.(2.1) and reconstructing $\sqrt{G}\phi$, as mentioned by [7], so called "C-property" needs to be preserved. Inspired by [11], we reconstruct $\sqrt{G}\phi_t$ instead of $\sqrt{G}\phi$, where $\phi_t = \phi + \phi_s$ is total geopotential height, and the reconstruction method is introduced in the next section. The momentum equations need to be modified as follow

$$\begin{cases} \dfrac{\partial \sqrt{G}\phi}{\partial t} + \dfrac{\partial \sqrt{G}\phi u}{\partial x} + \dfrac{\partial \sqrt{G}\phi v}{\partial y} = 0 \\ \dfrac{\partial \sqrt{G}\phi u}{\partial t} + \dfrac{\partial \sqrt{G}\left(\phi uu + \frac{1}{2}G^{11}\phi_t^2\right)}{\partial x} + \dfrac{\partial \sqrt{G}\left(\phi uv + \frac{1}{2}G^{12}\phi_t^2\right)}{\partial y} = \psi_M^1 + \psi_C^1 + \psi_B^1 \\ \dfrac{\partial \sqrt{G}\phi v}{\partial t} + \dfrac{\partial \sqrt{G}\left(\phi uv + \frac{1}{2}G^{21}\phi_t^2\right)}{\partial x} + \dfrac{\partial \sqrt{G}\left(\phi vv + \frac{1}{2}G^{22}\phi_t^2\right)}{\partial y} = \psi_M^2 + \psi_C^2 + \psi_B^2 \end{cases} \quad (2.13)$$

and the bottom topography influence term is now expressed as

$$\psi_B = \sqrt{G}\phi_s G^{ij}\frac{\partial \phi_t}{\partial x^j} = \sqrt{G}\phi_s \begin{pmatrix} G^{11}\dfrac{\partial \phi_t}{\partial x} + G^{12}\dfrac{\partial \phi_t}{\partial y} \\ G^{21}\dfrac{\partial \phi_t}{\partial x} + G^{22}\dfrac{\partial \phi_t}{\partial y} \end{pmatrix} \quad (2.14)$$

The reconstruction variables are $(\sqrt{G}\phi_t, \sqrt{G}\phi u, \sqrt{G}\phi v)$.

We write the governing equation set to vector form

$$\frac{\partial \boldsymbol{q}}{\partial t} + \frac{\partial \boldsymbol{F}(\boldsymbol{q})}{\partial x} + \frac{\partial \boldsymbol{G}(\boldsymbol{q})}{\partial y} = \boldsymbol{S}(\boldsymbol{q}) \quad (2.15)$$

$$\boldsymbol{q} = \begin{bmatrix} \sqrt{G}\phi \\ \sqrt{G}\phi u \\ \sqrt{G}\phi v \end{bmatrix}, \boldsymbol{F} = \begin{bmatrix} \sqrt{G}\phi u \\ \sqrt{G}\left(\phi uu + \frac{1}{2}G^{11}\phi_t^2\right) \\ \sqrt{G}\left(\phi uv + \frac{1}{2}G^{21}\phi_t^2\right) \end{bmatrix}, \boldsymbol{G} = \begin{bmatrix} \sqrt{G}\phi v \\ \sqrt{G}\left(\phi uv + \frac{1}{2}G^{12}\phi_t^2\right) \\ \sqrt{G}\left(\phi vv + \frac{1}{2}G^{22}\phi_t^2\right) \end{bmatrix}, \boldsymbol{S} = \begin{bmatrix} 0 \\ \psi_M^1 + \psi_C^1 + \psi_B^1 \\ \psi_M^2 + \psi_C^2 + \psi_B^2 \end{bmatrix} \quad (2.16)$$

## 3. Numerical Discretization

The finite volume method evaluates the temporal tendency of cell average by net flux, the flux across cell edges is able to be obtained by gaussian quadrature, we calculate the field value on gaussian quadrature point by spatial reconstruction and determine the flux value by Riemann solver. In this section, we introduce three two types of reconstruction methods, two-dimensional reconstruction by tensor product polynomial (TPP), and two-dimensional WENO based on tensor product polynomial (WENO2D). Reconstruction provides two values on each gaussian quadrature point (GQP), we use AUSM(Advection Upstream Splitting Method) [19][20] and LMARS (Low Mach number Approximate Riemann Solver)[5] schemes as Riemann solvers to determine the flux value, after that the flux across the edges between adjacent cells is obtained by linear gaussian quadrature on each edge.

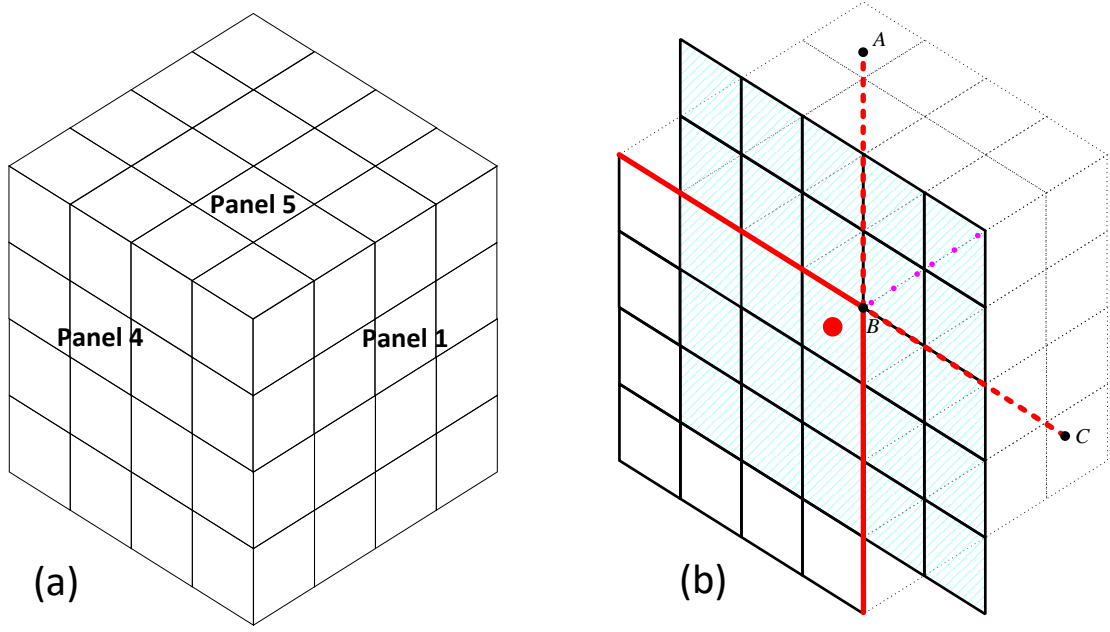

Figure 3.1 (a) Adjacent area of panels 1,4 and 5. (b) $5 \times 5$ reconstruction stencil nearby panel corner is represented by shade. The cell contains red dot is the target cell on panel 4, red solid lines are boundary of panel 4, red dash lines are extension line of panel 4 boundary line. $A$ and $C$ are points on dash line, $B$ is the upper right corner point of panel 4.

According to the finite volume scheme, average Eq.(2.15) on cell $i, j$, we have

$$\frac{\partial \overline{q}_{i,j}}{\partial t} + \frac{\overline{F}_{i+\frac{1}{2},j} - \overline{F}_{i+\frac{1}{2},j}}{\Delta x} + \frac{\overline{G}_{i+\frac{1}{2},j} - \overline{G}_{i+\frac{1}{2},j}}{\Delta y} = \overline{S}_{i,j} \tag{3.1}$$

$$\frac{\partial \overline{q}_{i,j}}{\partial t} = \frac{1}{\Delta x \Delta y} \frac{\partial}{\partial t} \iint_{\Omega_{i,j}} q \, dx \, dy, \quad \overline{S}_{i,j} = \frac{1}{\Delta x \Delta y} \iint_{\Omega_{i,j}} S \, dx \, dy \tag{3.2}$$

$$\overline{F}_{i-\frac{1}{2},j} = \frac{1}{\Delta y} \int_{e_{i-\frac{1}{2}}} F \, dy, \quad \overline{F}_{i+\frac{1}{2},j} = \frac{1}{\Delta y} \int_{e_{i+\frac{1}{2}}} F \, dy \tag{3.3}$$

$$\overline{G}_{i,j-\frac{1}{2}} = \frac{1}{\Delta x} \int_{e_{j-\frac{1}{2}}} G \, dx, \quad \overline{G}_{i,j+\frac{1}{2}} = \frac{1}{\Delta x} \int_{e_{j+\frac{1}{2}}} G \, dx \tag{3.4}$$

where $\Omega_{i,j}$ represents the region overlapped by cell $(i,j)$, $e_{i-\frac{1}{2}}, e_{i+\frac{1}{2}}, e_{j-\frac{1}{2}}, e_{j+\frac{1}{2}}$ are left, right, bottom, top edges of cell $(i,j)$.

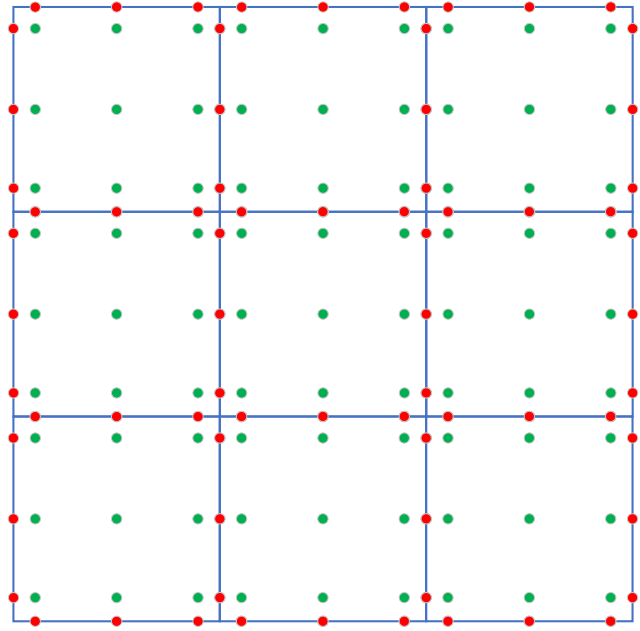

Figure 3.2 Function points on cell. Red points are edge quadrature points (EQP) or called flux points, green points are inner cell quadrature points (CQP).

The physical interpretation of equation Eq.(3.1) is that the average tendency of prognostic field $q$ within cell $(i,j)$ is governed by the average net flux and average source. In this study, we calculate these averages using Gaussian quadrature, the function points within each cell are illustrated in Figure 3.2, the EQPs are share by adjacent cells, and CQPs are exclusive for each cell.

Average on edge by 1D scheme:

$$\overline{F}_{i+\frac{1}{2},j} \approx \sum_{r=1}^{m_e} w_r F_r = w F_r \tag{3.5}$$

where $w = (w_1, w_2, \dots, w_{m_e})$ is the 1D Gaussian quadrature coefficient matrix, $m_e$ is the number of quadrature points on each edge.

Average in cell by 2D scheme:

$$\overline{S}_{i,j} \approx \sum_{r=1}^{m_c} W_r S_r = W S_r \tag{3.6}$$

where $W = (W_1, W_2, \dots, W_{m_c})$ is the 2D Gaussian quadrature coefficient matrix, $m_e$ is the number of quadrature points on each cell.

## 3.1 Tensor Product Polynomial (TPP) Reconstruction

The computational space of cubed sphere is rectangular and structured, we adopt to take reconstruction on square stencil. A two-dimensional $d$-th degree polynomial has number of terms $n = \frac{(d+1)(d+2)}{2}$, it is not able to be fully filled by a $k$-th order square stencil ($k \times k$ cells), as shown in Figure 3.3 (a).

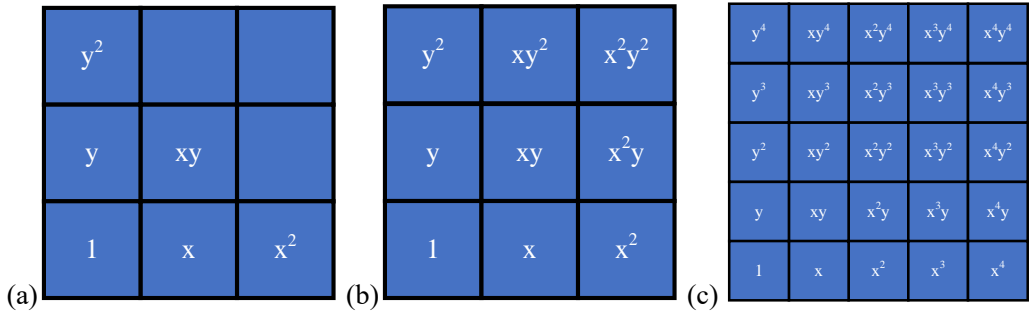

Figure 3.3 Polynomial terms on stencils. (a): 2nd degree polynomial stencil; (b): 3rd order TPP stencil; (c) 5th order TPP stencil

We make use of the TPP to approximate the horizontal reconstruction. A TPP is expressed as

$$p(x,y) = \sum_{i=1}^{m}\sum_{j=1}^{n} a_k x^{i-1} y^{j-1} = \sum_{k=1}^{N} a_k c_k(x,y) \tag{3.7}$$

where $m$ and $n$ are row and column of stencil. $a_k$ is the coefficient of each term, the term index $k = i + m(j-1)$, and $c_k(x,y) = x^\alpha y^\beta$, $\alpha = k - int\left(\frac{k-1}{n}\right)n$, $\beta = int\left(\frac{k-1}{n}\right)$, int is the same function in Fortran, $N$ is the cell number in stencil and also the term number of the TPP, the 3rd and 5th order stencils are shown in Figure 3.3. We define column vectors $c(x,y) = \{c_k(x,y) | k = 1,2,3,\ldots,N\}$ and $\boldsymbol{a} = \{a_k | k = 1,2,3,\ldots,N\}$, the point value on $(x,y)$ can be written as

$$p(x,y) = \boldsymbol{c}(x,y) \cdot \boldsymbol{a} \tag{3.8}$$

The volume integration average (VIA) of evolution field $q$ on cell $\Omega_i$ is represented by

$$\bar{q}_i = \frac{1}{\Delta x_i \Delta y_i} \iint_{\Omega_i} p(x,y) dx dy \tag{3.9}$$

$\Delta x_i, \Delta y_i$ are length of edges $x, y$ of cell $\Omega_i$ in computational space. In our setting, all of the cells in the computational space are set to unit square, therefore $\Delta x_i = 1, \Delta y_i = 1$, and (3.9) becomes

$$\bar{q}_i = \iint_{\Omega_i} p(x,y) dx dy = \iint_{\Omega_i} \boldsymbol{c} \cdot \boldsymbol{a}\, dx dy = \boldsymbol{\psi}_i \cdot \boldsymbol{a} \tag{3.10}$$

where $\boldsymbol{\psi}_i = \iint_{\Omega_i} \boldsymbol{c}\, dx dy = \begin{pmatrix} \iint_{\Omega_i} c_1 dx dy \\ \iint_{\Omega_i} c_2 dx dy \\ \vdots \\ \iint_{\Omega_i} c_N dx dy \end{pmatrix}$, combining $N$ cells, we have following linear system

$$A\boldsymbol{a} = \bar{\boldsymbol{q}} \tag{3.11}$$

$$A = \begin{pmatrix} \boldsymbol{\psi}_1^T \\ \boldsymbol{\psi}_2^T \\ \vdots \\ \boldsymbol{\psi}_N^T \end{pmatrix}, \bar{\boldsymbol{q}} = \begin{pmatrix} \bar{q}_1 \\ \bar{q}_2 \\ \vdots \\ \bar{q}_N \end{pmatrix} \tag{3.12}$$

and polynomial coefficient $\boldsymbol{a}$ can be obtain by solving Eq.(3.11).

$$\boldsymbol{a} = A^{-1}\bar{\boldsymbol{q}} \tag{3.13}$$

The reconstruction values on $M$ points can be obtained by following formula

$$P = C\boldsymbol{a} = CA^{-1}\bar{\boldsymbol{q}} = R\bar{\boldsymbol{q}} \tag{3.14}$$

where $P = \begin{pmatrix} p(x_1,y_1) \\ p(x_2,y_2) \\ \vdots \\ p(x_M,y_M) \end{pmatrix}, C = \begin{pmatrix} \boldsymbol{c}_1^T \\ \boldsymbol{c}_2^T \\ \vdots \\ \boldsymbol{c}_M^T \end{pmatrix}, \boldsymbol{c}_j^T = \boldsymbol{c}^T(x_j,y_j), j = 1,2,\ldots,M$, superscript $T$ stands for transpose matrix, and

the reconstruction matrix

$$R = CA^{-1} \tag{3.15}$$

In our model, $(x_j, y_j)$ represents the function points on target cell.

## 3.2 Genuine Two-Dimensional WENO

WENO is an adaptive algorithm, it takes high order accuracy in smooth area, and when the field is discontinuous, WENO reduce the accuracy to low order to capture the shock. Shi and Shu (2002)[33] mentioned a fifth-order finite volume WENO can be constructed in two different ways, "Genuine 2D" and "Dimension by Dimension", in genuine 2D case, a $3^{rd}$ order stencil with $3 \times 3$ cells can be decomposed by sub-stencils with $2 \times 2$ cells, and a $5 \times 5$ stencil can be decomposed to 9 sub-stencils, there are $3 \times 3$ cells contained in each sub-stencil, details in Figure 3.4 and Figure 3.5. The crucial issues are determining the optimal linear weight in two-dimensional stencil. Once the optimal linear weights are determined, the combination of sub-stencils provides $5^{th}$ order accuracy in smooth field. Authors of [33] mentioned the linear weight can be calculated by Lagrange interpolation basis, but no more details are provided. In this section, we introduce the method of constructing WENO 2D with $3^{rd}$ and $5^{th}$ order by least square method.

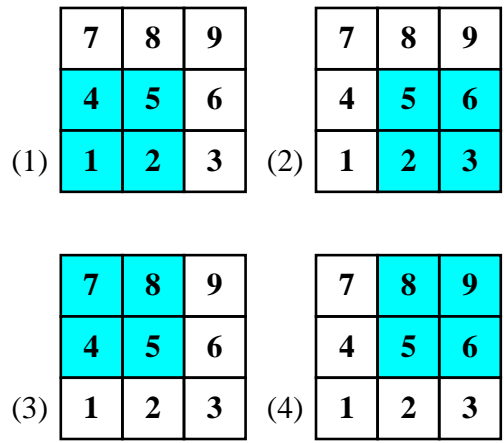

Figure 3.4 Stencils of $3^{rd}$ order WENO 2D. The high order stencil contains cells 1~9, blue ones represent the cells in sub-stencils (1) ~ (4).

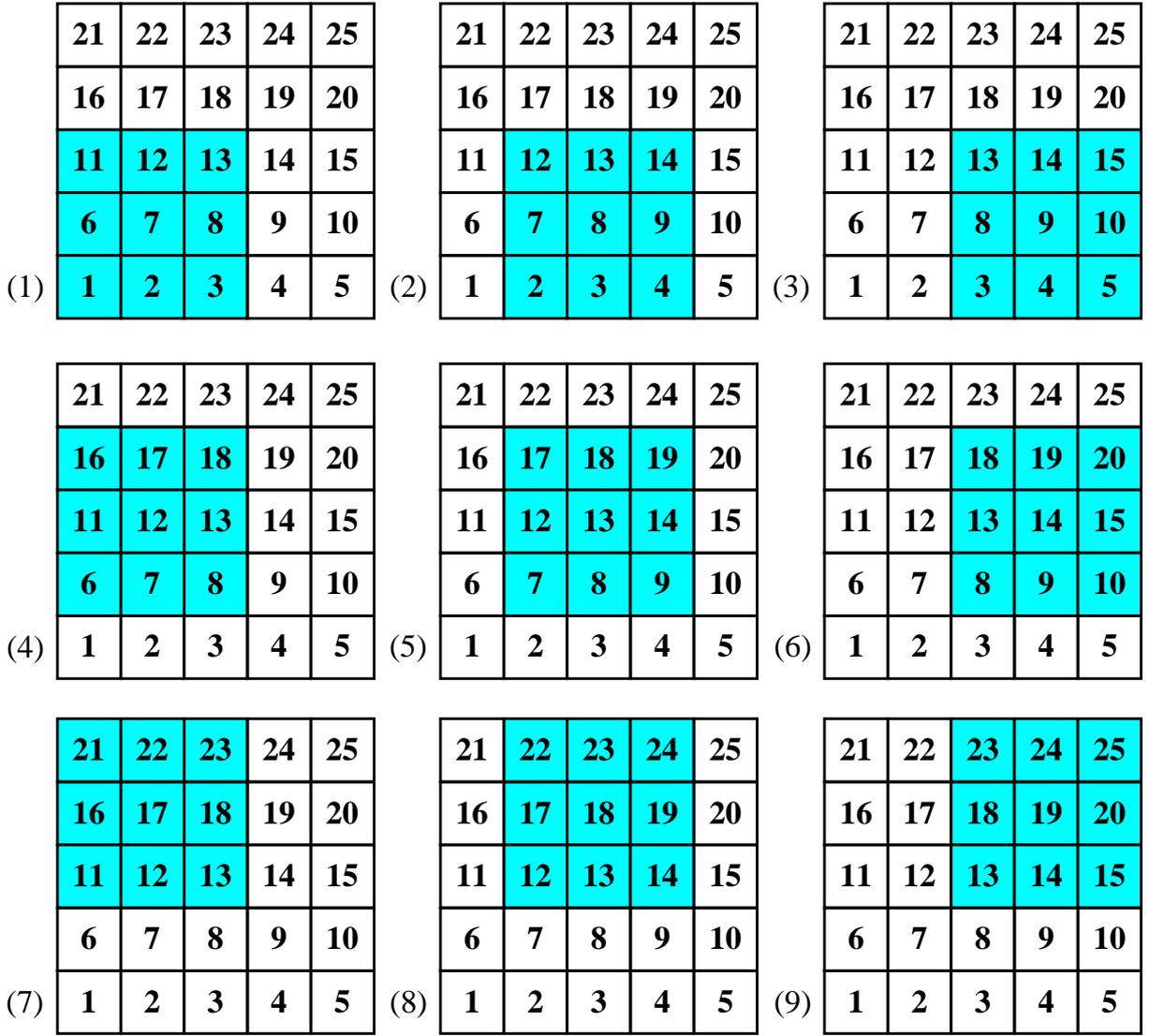

Figure 3.5 Stencils of 5th order WENO 2D. The high order stencil contains cells 1~25, blue ones represent the cells in sub-stencils (1) ~ (9).

We construct WENO 2D based on TPP and square stencil. As mentioned in previous section, a $n$-th order stencil contains $m = n^2$ cells, and the stencil width is $h = n$. Decomposing the high-order stencil into $s = \left(\frac{n+1}{2}\right)^2$ sub-stencils, there are $s_c = s$ cells in each sub-stencil, and the sub-stencil width is $l = \frac{n+1}{2}$. For the reconstruction point $(x, y)$, suppose $p_H(x, y)$ is the reconstruction value of high-order stencil, the reconstruction values of sub-stencils are stored in vector $\boldsymbol{p} = \left(p_1(x,y), p_2(x,y), \cdots, p_s(x,y)\right)^T$. The intention of constructing the optimal linear weights is to determine the unique weights $\gamma = (\gamma_1, \gamma_2, \cdots, \gamma_s)$, such that

$$p_H = \gamma \boldsymbol{p} \tag{3.16}$$

For calculating $\gamma$, we need to write the high order and low order reconstruction matrix into the same linear system. For sub-stencil $i$ we have reconstruction matrix $R_i = (r_{ik}), k = 1,2,\ldots,s_c$, which is computed by (3.15). We define $R_{L_i} = \left(r_{L_{ij}}\right), j = 1,2,\ldots,m$ is the extension matrix of $R_i$, and

$$(R_i)_{1 \times s_c} (E)_{s_c \times m} = (R_{L_i})_{1 \times m}$$

subscript outside bracket represents the shape of each matrix in bracket, and the matrix $E = (e_{ij}), i = 1,2,\ldots, s_c; j = 1,2,\ldots, m$ describes the correspondence between cells in high-order stencil and low-order stencil, when the $i$-th cell in low-order stencil is the same as the $j$-th cell in high order stencil, $e_{ij} = 1$, otherwise, $e_{ij} = 0$. The example case of the 3$^{rd}$ order scheme is shown in section 7, the high order situations are similar to the 3$^{rd}$ order case.

Expand (3.16) by (3.14) in single point case ($M = 1$), yield

$$R_H \bar{q} = \sum_{i=1}^{s} R_{L_i} \gamma_i \bar{q} \tag{3.17}$$

where the elements of vector $\bar{q} = (q_1, q_2, \cdots, q_m)^T$ represent VIA of each cell in high-order stencil. $R_H = (r_{H_j}), j = 1,2,\ldots, m$ is the reconstruction matrix of high-order stencil.

We set $R_L = (R_{L_1}, R_{L_2}, \ldots, R_{L_s})^T$, (3.17) becomes

$$R_L \gamma = R_H \tag{3.18}$$

the unknown optimal weight matrix $\gamma$ can be determined by following least square procedure

$$\gamma = (R_L^T R_L)^{-1} R_L^T R_H \tag{3.19}$$

However, the elements of $\gamma$ could be negative, which would cause unstable. A split technique mentioned by Shi et al. (2002)[33] was adopted to solve this problem. The optimal weights can be split into two parts:

$$\gamma^+ = \frac{\theta |\gamma| + \gamma}{2}, \quad \gamma^- = \frac{\theta |\gamma| - \gamma}{2} \tag{3.20}$$

where the constant $\theta = 3$. The reconstruction value on point $(x, y)$:

$$q(x, y) = \sum_{i=1}^{s} (\omega_i^+ - \omega_i^-) p_i(x, y) \tag{3.21}$$

We want the nonlinear weight $\omega_i$ is large when stencil $i$ is smooth on target cell and if stencil $i$ is discontinuous, $\omega_i$ should be a small value. There are serial choices of nonlinear weight scheme WENO-JS[12], WENO-Z[4], WENO-Z+[1], WENO-Z+M[23] and so on. In this paper, we adopt WENO-Z scheme, most of WENO schemes are developed based on one-dimensional case, we extend WENO-Z to a two-dimensional case by modifying $\tau$, which is an important coefficient for high order WENO-Z. For stencil $i$ the nonlinear weight is given as

$$\omega_i^\pm = \frac{\alpha_i^\pm}{\sum_{i=1}^{s} \alpha_i^\pm} \tag{3.22}$$

$$\alpha_i^\pm = \gamma_i^\pm \left(1 + \frac{\tau}{\beta_i + \varepsilon}\right) \tag{3.23}$$

$$\tau = \frac{2}{(s+1)s} \sum_{\eta=1}^{s-1} \sum_{\psi=\eta}^{s} |\beta_\psi - \beta_\eta| \tag{3.24}$$

The smooth indicators $\beta_i$ measure how smooth the reconstruction functions are in the target cell; we use a similar scheme as described in [49]:

$$\beta_j = \sum_{\zeta=1}^{m} \iint_\Omega \frac{\partial^\zeta}{\partial x^{\zeta_1} \partial y^{\zeta_2}} p_j(x,y) dx dy \qquad (3.25)$$

where $\zeta_1 + \zeta_2 = \zeta$ and $\zeta > 0$, $\zeta_1, \zeta_2 \in [0, n]$.

## 3.3 Treatment of the Panel Boundaries

The cubed sphere grid comprises eight panel boundaries, and the flux across the interface between any two panels must be computed at the quadrature points situated on the edges of the boundary cells, as depicted in Figure 3.6 (a). However, a challenge arises because the coordinates across these panel boundaries are discontinuous. Given that the TPP reconstruction necessitates a square stencil, the values of the cells outside the domain (referred to as ghost cells) must be computed through interpolation within the adjacent panel, as illustrated in Figure 3.6 (b). Ullrich et al. (2010) [37] proposed a one-side interpolation scheme, but in our test, we found that using one-sided interpolation around panel boundaries leads to instability when the accuracy exceeds the 7th order.

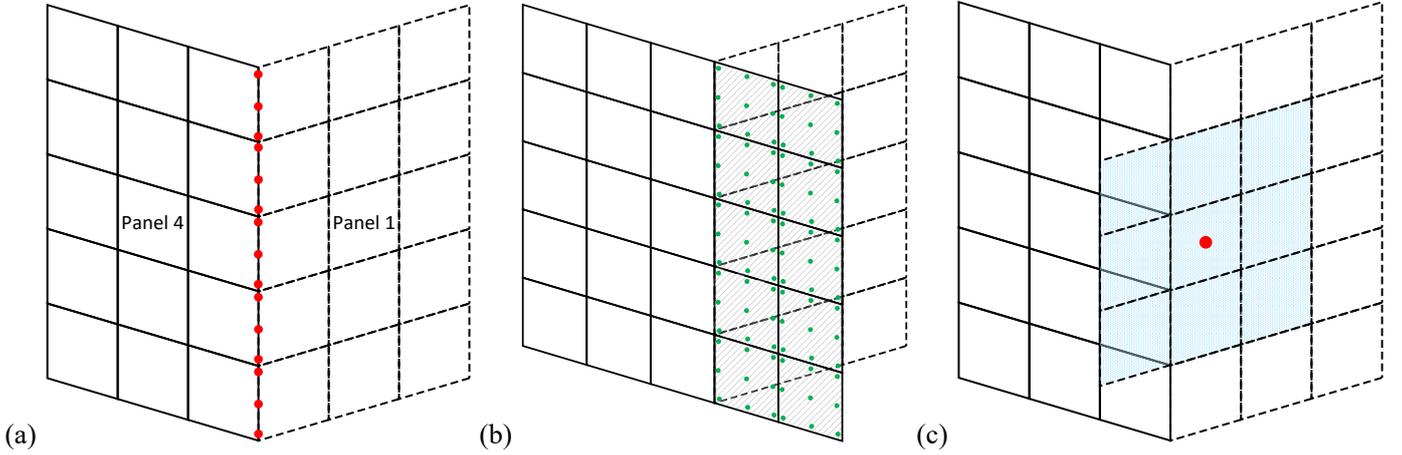

Figure 3.6 Points and cells close to panel boundary. (a) Flux points on the interface between Panel 1 and Panel 4, the flux across each panel at these points are determined by Riemann solver, which merges the reconstruction outcomes from both panels into a single flux value; (b) Ghost cells (shaded cells) out of Panel 4 boundary, with green points representing the GQP in these cells; (c) Cells requirement for 3rd order ghost cell interpolation stencil, red point represents arbitrary GQP which is in the ghost cell on Panel 4, the blue shaded region represents the TPP reconstruction stencil (on Panel 1) to interpolate this red GQP.

### 3.3.1 Ghost Cell Interpolation

To achieve arbitrary high order accuracy, we attempt to devise a ghost cell interpolation scheme that incorporates information from both sides of the panel boundary. It's clear that the ghost cell values are unknown prior to interpolation, our preliminary idea is estimating the ghost cell values through an iterative process. Specifically, the method entails repeatedly performing the ghost cell interpolation until the increments of the cell values converge to zero.

Through mathematical analysis, we found that this iterative process is able to be express as a linear mapping, the iteration is no longer necessary, the detailed derivation is provided in the appendix. However, obtaining the mapping matrix of the interpolation process, we have to compute a large inverse matrix, which is not only computationally expensive, but also incurs too large memory requirements. To overcome this challenge, we implement the iterative interpolation process in PyTorch code, and leverage its automatic differentiation capability to directly obtain the interpolation matrix.

With reference to the appendix, we introduce the process of this method. Firstly, we initialize all of the ghost cell values to zero, denoted as $\boldsymbol{g}^{(0)} = 0$, the superscript indicates the iteration number. Secondly, interpolating the GQP in the ghost cells by TPP stencil. To illustrate, consider two adjacent panels shown in Figure 3.6(a). For any out-domain cell in panel 4 (shaded cell in Figure 3.6(b)), the GQPs in the cell are actually locating in panel 1. We interpolate the GQPs using the TPP stencil depicted in Figure 3.6(c), which includes some ghost cells of panel 1. After interpolating all of the GQPs, the ghost cell values using the Gaussian quadrature Eq.(3.6), obtaining $\boldsymbol{g}^{(1)}$. We then compute the norm 2 residual $r^{(k)} = \|\boldsymbol{g}^{(k+1)} - \boldsymbol{g}^{(k)}\|_2$. Repeat the second step until $r^{(k)} < \epsilon$, with $\epsilon = 1.e^{-14}$ for double precision and $\epsilon = 1.e^{-5}$ for single precision. In practical applications, we have observed that the iteration typically converges within fewer than 10 loops, hence we set the loop count to 10 for consistency. After this stage, we have obtained the mapping from known cells to ghost cells $G: \boldsymbol{q} \to \boldsymbol{g}$. According to Eq.(7.12) in appendix, this mapping in linear, implying $G = \frac{\partial \boldsymbol{g}}{\partial \boldsymbol{q}}$ is a linear matrix, we can easily compute this derivative by using "autograd" function in PyTorch.

### 3.3.2 Fields Conversion Between Panels

The coordinates on panels are different. To explain the method of conversion fields between panels, we provide an example between panel 1 and 4. As shown in Figure 3.6(a), the flux points are shared by two panels, the coordinates are discontinuous on the panel interface. Consequently, we must reset the metric on mass variable, and the coordinate of wind vectors are also need to be converted from one panel to the other.

Suppose $\boldsymbol{q}_1 = \left[(\sqrt{G}\phi)_1, (\sqrt{G}\phi u)_1, (\sqrt{G}\phi v)_1\right]^T$ and $\boldsymbol{q}_4 = \left[(\sqrt{G}\phi)_4, (\sqrt{G}\phi u)_4, (\sqrt{G}\phi v)_4\right]^T$ represent the fields on panel 1 and 4. The mass field conversion from panel 4 to panel 1 is expressed by

$$(\sqrt{G}\phi)_4^1 = \frac{\sqrt{G}_4}{\sqrt{G}_1}(\sqrt{G}\phi)_1 \tag{3.26}$$

the subscript represents the target panel and the superscript stands for source panel.

The momentum vector is converted by two steps. Firstly, we convert the contravariant momentum from panel 1 to spherical momentum by matrix $J$ as we mentioned in Eq.(2.10), then convert spherical momentum to contravariant momentum in panel 4.

$$\begin{pmatrix}(\sqrt{G}\phi u_s)_1 \\ (\sqrt{G}\phi v_s)_1\end{pmatrix} = J_1 \begin{pmatrix}(\sqrt{G}\phi u)_1 \\ (\sqrt{G}\phi v)_1\end{pmatrix} \tag{3.27}$$

$$\begin{pmatrix}(\sqrt{G}\phi u)_4 \\ (\sqrt{G}\phi v)_4\end{pmatrix} = J_4^{-1} \frac{\sqrt{G}_4}{\sqrt{G}_1}\begin{pmatrix}(\sqrt{G}\phi u_s)_1 \\ (\sqrt{G}\phi v_s)_1\end{pmatrix} \tag{3.28}$$

where $J_1$ is the $J$ matrix on panel 1, $J_4^{-1}$ is the inverse matrix of $J$ on panel 4. Obviously, the vector conversion is linear,

therefore Eq.(3.27) and Eq.(3.28) can be merged into following equation

$$\begin{pmatrix} (\sqrt{G}\phi u)_4 \\ (\sqrt{G}\phi v)_4 \end{pmatrix} = C \begin{pmatrix} (\sqrt{G}\phi u)_1 \\ (\sqrt{G}\phi v)_1 \end{pmatrix} \tag{3.29}$$

where matrix $C = \frac{\sqrt{G_4}}{\sqrt{G_1}} J_4^{-1} J_1$.

The mass and vector are also need to be converted on GQPs in the same manner.

## 3.4 Riemann Solver

After performing spatial reconstruction, two distinct reconstruction outcomes emerge on either side of a given point location, as noted by Chen et al. (2013) [5], since the majority of atmospheric flow speeds correspond to small Mach numbers, we adopt the Low Mach number Approximate Riemann Solver (LMARS) as Riemann solver to determine the flux at the edge quadrature points.

Spatial reconstruction gives the left and right state vector,

$$q_L = \begin{bmatrix} (\sqrt{G}\phi)_L \\ (\sqrt{G}\phi u)_L \\ (\sqrt{G}\phi v)_L \end{bmatrix}, \quad q_R = \begin{bmatrix} (\sqrt{G}\phi)_R \\ (\sqrt{G}\phi u)_R \\ (\sqrt{G}\phi v)_R \end{bmatrix} \tag{3.30}$$

First of all, we convert the contravariant wind $u$ to physical speed $u^\perp$ that is perpendicular to the cell edge.

$$u^\perp = \frac{u}{\sqrt{G^{ii}}}, \quad i = \begin{cases} 1, & x\ direction \\ 2, & y\ direction \end{cases} \tag{3.31}$$

The wind speed $m^*$ and geopotential height $\phi$ are calculated by

$$m^* = \frac{1}{2}\left(u_L^\perp + u_R^\perp - \frac{\phi_R - \phi_L}{c}\right) \tag{3.32}$$

$$\phi = \frac{1}{2}[\phi_L + \phi_R - c(u_R^\perp - u_L^\perp)] \tag{3.33}$$

$$c = \frac{c_L + c_R}{2} \tag{3.34}$$

$$c_L = \sqrt{\phi_L}, c_R = \sqrt{\phi_R} \tag{3.35}$$

$c$ is the phase speed of external gravity wave, the subscript $L, R$ represent the left and right side of cell edge.

Once $m^*$ is determined, we need to convert it back to contravariant speed by

$$m = m^* \sqrt{G^{ii}} \tag{3.36}$$

The flux across the cell edge is then given by

$$F = \frac{1}{2}[m(q_L + q_R) - sign(m)(q_R - q_L)] + P \tag{3.37}$$

$$P = \begin{pmatrix} 0 \\ \frac{1}{2}\sqrt{G}G^{1i}\phi_t^2 \\ \frac{1}{2}\sqrt{G}G^{2i}\phi_t^2 \end{pmatrix}, \quad i = \begin{cases} 1, & x\ direction \\ 2, & y\ direction \end{cases} \tag{3.38}$$

For calculation of $H$ the method is similar.

## 3.5 Temporal Integration

We use the explicit Runge-Kutta (RK) as time marching scheme, Wicker and Skamarock (2002) described a 3rd order RK with three stages[40], for the prognostic fields $q$, the integration step from time slot $n$ to $n+1$:

$$q^* = q^n + \frac{\Delta t}{3}\left(\frac{\partial q^n}{\partial t}\right) \quad (3.39)$$

$$q^{**} = q^* + \frac{\Delta t}{2}\left(\frac{\partial q^*}{\partial t}\right) \quad (3.40)$$

$$q^{n+1} = q^n + \Delta t\left(\frac{\partial q^{**}}{\partial t}\right) \quad (3.41)$$

where $\Delta t$ is the time step, and temporal tendency terms $\frac{\partial q}{\partial t}$ can be obtain by (2.15), (2.16).

# 4. High Performance Implementation and Automatic Differentiation

The spatial operator and temporal integration of HOPE can be easily implemented using PyTorch. Specifically, the spatial reconstruction given by Eq.(3.14) is analogous to a convolution operation, while the Gaussian quadrature can be efficiently expressed as a matrix-vector multiplication. Both of these operations are highly optimized for execution on GPUs, ensuring superior performance. Furthermore, as a versatile platform for AI development, PyTorch offers automatic differentiation capabilities for all the aforementioned functions, streamlining the implementation and enabling efficient gradient computation.

For the reconstruction implementation. Suppose the cubed sphere grid comprises $n_c$ cells in $x$-direction within each panel, including ghost cells. The panel number is $n_p$, and the shallow water equation involves $n_v$ prognostic variables, we write the cell state tensor $q$ with the shape $(n_v n_p, 1, n_c, n_c)$. The TPP reconstruction weight tensor $R$ has shape $(n_{poc}, 1, s_w, s_w)$, where $n_{poc}$ is the number of points required to be interpolated within each cell (including EQP and CQP), $s_w$ denotes the stencil width. The reconstruction can be executed by a simple command (pseudo-code):

$$q_{rec} = torch.nn.Functional.conv2d(q, R) \quad (4.1)$$

where the shape of $q_{rec}$ is $(n_v n_p, n_{poc}, n_c, n_c)$

For the Gaussian quadrature implementation. Suppose the edge state tensor $q_e$ with the shape $(n_v, n_p, n_c, n_c, n_{poe})$, where $n_{poe}$ is the number of quadrature points on each edge. The edge Gaussian quadrature weight tensor $g_e$ has shape $(n_{poe})$. The quadrature is expressed by:

$$q_g = torch.matmul(q_e, g_e) \quad (4.2)$$

where the shape of $q_g$ is $(n_v, n_p, n_c, n_c)$

After spatial reconstruction, the resulting data is utilized in the Riemann solver for EQPs and for source term computation on CQPs. Subsequently, integration is performed on both EQPs and CQPs to calculate the net flux and the cell-averaged source term tendency. However, there is a dimensionality mismatch between the reconstructed points, i.e. $n_{poc}$ is the first dimension of $q_{rec}$, while $n_{poe}$ is the last dimension of $q_e$. To address this dimensionality issue, two methods are available. The first method involves rearranging the $n_{poc}$ dimension to the last position using the "torch.tensor.permute" operation in PyTorch, This allows Gaussian integrations to be naturally implemented through the "torch.matmul" operation. The second method, which avoids the need for the "permute" operation, maintains the original dimension sequence. Instead, Gaussian integrations are performed using the "torch.einsum" function. This function sums the product of the elements of the input arrays along dimensions specified using a notation based on the Einstein summation convention.

$$q_g = torch.\text{einsum}('vnpij, p \to vnij', q_e, g_e) \tag{4.3}$$

We have conducted performance tests comparing the two methods, and the results indicate that the second method is approximately 5% faster than the first. Specifically, the first method took 649 seconds, while the second method took 616 seconds. The test was set as a one-day steady state geostrophic flow (with details provided in section 5.1) simulation at a resolution of 0.1°, using 3rd order accuracy reconstruction stencil. The time step was 30 seconds, and the default data type used was "torch.float32" (single precision).

The Riemann solver implementation on flux points is way easier, only needs to call "torch.sign" for Eq.(3.37), while all other operations can be executed using basic arithmetic: addition, subtraction, multiplication, and division. During a Runge-Kutta sub-step, there are no dependencies, and neither "for" loops nor "if" statements are required in the HOPE kernel code. This algorithm fully leverages the capabilities of the GPU.

## 5. Numerical Experiments

The standard test cases for spherical shallow water model are mentioned by Williamson et al. (1992)[41]. In this article, we test HOPE dynamic core using case number 2, 5, 6 desired in [41], and the case of perturbed jet flow mentioned in [8]. Besides, we have designed a dam-break experiment to prove the ability of the HOPE model in capturing shock waves.

In our experiments, the grid resolutions are denoted by the count of cells along one dimension on each panel of the cubed sphere; for instance, C90 signifies that each panel is subdivided into a $90 \times 90$ grid, corresponding to a grid interval of $\Delta x = \Delta y = 1°$.

### 5.1 Steady State Geostrophic Flow

Steady state geostrophic flow is the 2nd case in Williamson et al. (1992)[41], it provided an analytical solution for spherical shallow water equations, it was wildly used in accuracy test for shallow water models. The analytical solution is a steady state, which means the initial filed is the exact solution. The initial field is expressed as

$$\phi = \phi_0 - \left(a\Omega u_0 + \frac{u_0^2}{2}\right)(-\cos\lambda \cos\theta \sin\alpha + \sin\theta \cos\alpha)^2 \tag{5.1}$$

$$u_s = u_0(\cos\theta \cos\alpha + \cos\lambda \sin\theta \sin\alpha) \tag{5.2}$$

$$v_s = -u_0 \sin \lambda \sin \alpha \tag{5.3}$$

where $\lambda, \theta$ are longitude and latitude, $\phi$ is geopotential height, $u_s, v_s$ are zonal wind and meridional wind, earth radius is $a = 6371220\ m$, earth rotation angular velocity $\Omega = 7.292 \times 10^{-5}\ s^{-1}$, basic flow speed $u_0 = \frac{2\pi a}{12*86400}\ m/s$, basic geopotential height $\phi_0 = 29400\ m^2/s^2$, $\alpha = 0$ and gravity acceleration $g = 9.80616\ m/s^2$. The conversion between the spherical wind $(u_s, v_s)$ and contravariant wind is given by (2.9).

We use three kinds of norm errors to measure the simulation errors,

$$L_1 = \frac{I[\phi(i,j,p) - \phi_{ref}(i,j,p)]}{I[\phi_{ref}(i,j,p)]} \tag{5.4}$$

$$L_2 = \sqrt{\frac{I\left[\left(\phi(i,j,p) - \phi_{ref}(i,j,p)\right)^2\right]}{I[\phi_{ref}^2(x,y,p)]}} \tag{5.5}$$

$$L_\infty = \frac{\max|\phi(i,j,p) - \phi_{ref}(i,j,p)|}{\max|\phi_{ref}(i,j,p)|} \tag{5.6}$$

$$I(\phi) = \sum_{p=1}^{n_p} \sum_{j=1}^{n_y} \sum_{i=1}^{n_x} (\sqrt{G}\phi)_{ijp} \tag{5.7}$$

where $n_x, n_y$ represent the number cells in $x, y$ directions, and $n_p = 6$ is the number of panels on cubed sphere grid. The metric Jacobian $\sqrt{G}$ has the same definition as Eq.(2.8). For example, a C90 grid corresponds $n_x = n_y = 90$.

We simulated the steady state geostrophic flow over one period (12 days) to test the norm errors and corresponds convergence rate. Since the norm error becomes too small to express by double precision number, all of the experiments were based on the quadruple precision version of HOPE. Time steps were set to $\Delta t = 600, 400, 200, 100, 50\ s$ for C30, C45, C90, C180 and C360, respectively.

Table 5.1 Norm errors and convergence rates of steady state geostrophic flow at day 12.

| TPP3 | C30 | C45 | C90 | C180 | C360 |
|---|---|---|---|---|---|
| $L_1$ error | 1.8853E-03 | 5.6474E-04 | 7.0960E-05 | 8.8777E-06 | 1.1099E-06 |
| $L_1$ rate |  | 2.9731 | 2.9925 | 2.9988 | 2.9998 |
| $L_2$ error | 2.1484E-03 | 6.4171E-04 | 8.0500E-05 | 1.0069E-05 | 1.2588E-06 |
| $L_2$ rate |  | 2.9802 | 2.9949 | 2.9991 | 2.9998 |
| $L_\infty$ error | 4.3242E-03 | 1.2932E-03 | 1.6201E-04 | 2.0275E-05 | 2.5350E-06 |
| $L_\infty$ rate |  | 2.9770 | 2.9968 | 2.9983 | 2.9997 |
| **TPP5** |  |  |  |  |  |
| $L_1$ error | 3.6122E-06 | 4.7493E-07 | 1.4827E-08 | 4.6322E-10 | 1.4474E-11 |
| $L_1$ rate |  | 5.0039 | 5.0014 | 5.0004 | 5.0002 |
| $L_2$ error | 5.2427E-06 | 6.9169E-07 | 2.1627E-08 | 6.7584E-10 | 2.1119E-11 |
| $L_2$ rate |  | 4.9954 | 4.9992 | 5.0000 | 5.0001 |
| $L_\infty$ error | 1.6810E-05 | 2.2451E-06 | 7.0534E-08 | 2.2070E-09 | 6.8985E-11 |
| $L_\infty$ rate |  | 4.9652 | 4.9923 | 4.9982 | 4.9996 |
| **TPP7** |  |  |  |  |  |
| $L_1$ error | 8.1697E-08 | 4.7967E-09 | 3.7678E-11 | 2.9547E-13 | 2.3125E-15 |
| $L_1$ rate |  | 6.9922 | 6.9922 | 6.9946 | 6.9974 |
| $L_2$ error | 8.7991E-08 | 5.1644E-09 | 4.0507E-11 | 3.1728E-13 | 2.4823E-15 |
| $L_2$ rate |  | 6.9931 | 6.9943 | 6.9963 | 6.9979 |
| $L_\infty$ error | 1.4741E-07 | 8.6376E-09 | 6.7814E-11 | 5.3387E-13 | 4.1901E-15 |
| $L_\infty$ rate | | 6.9971 | 6.9929 | 6.9889 | 6.9934 |
| **TPP9** | | | | | |
| $L_1$ error | 7.8909E-10 | 2.1780E-11 | 4.3925E-14 | 8.6359E-17 | |
| $L_1$ rate | | 8.8537 | 8.9538 | 8.9905 | |
| $L_2$ error | 9.5638E-10 | 2.6409E-11 | 5.3341E-14 | 1.0494E-16 | |
| $L_2$ rate | | 8.8526 | 8.9516 | 8.9896 | |
| $L_\infty$ error | 2.3946E-09 | 6.6773E-11 | 1.3547E-13 | 2.6644E-16 | |
| $L_\infty$ rate | | 8.8285 | 8.9452 | 8.9899 | |
| **TPP11** | | | | | |
| $L_1$ error | 1.1908E-10 | 1.3799E-12 | 6.7696E-16 | 3.3197E-19 | |
| $L_1$ rate | | 10.9943 | 10.9932 | 10.9938 | |
| $L_2$ error | 1.3084E-10 | 1.5186E-12 | 7.4489E-16 | 3.6500E-19 | |
| $L_2$ rate | | 10.9904 | 10.9934 | 10.9949 | |
| $L_\infty$ error | 2.4204E-10 | 2.8579E-12 | 1.4147E-15 | 6.9567E-19 | |
| $L_\infty$ rate | | 10.9479 | 10.9803 | 10.9898 | |

In Table 5.1, we present the geopotential height simulation errors and convergence accuracy of different order accuracy schemes at various resolutions. It is evident that HOPE is capable of achieving the designed accuracies in all tests. When the resolution exceeds C180, the errors obtained from the 7th, 9th, and 11th-order precision schemes have surpassed the limits expressible by double-precision numbers. This demonstrates HOPE's excellent error convergence for simulating smooth flow fields. It should be noted that high-order accuracy schemes do consume more computational resources. HOPE has proven the feasibility of ultra-high-order accuracy finite volume methods on cubed sphere grids. However, in simulating the real atmosphere, a balance between computational efficiency and error must be considered. We believe that 3rd or 5th order accuracy schemes will be more practical for subsequent developments in baroclinic atmosphere model.

## 5.2 Zonal Flow over an Isolated Mountain

Zonal flow over an isolated mountain is the 5th case mentioned in Williamson et al. (1992)[41], this case was usually be implemented to test the topography influence in shallow water models. The initial condition is defined by Eq.(5.1)~(5.3), but $h_0 = 5960\ m, \phi_0 = h_0 g, u_0 = 20 m/s$. The mountain height is expressed as

$$h_s = h_{s0}\left(1 - \frac{r}{R}\right) \tag{5.8}$$

where $h_{s0} = 2000\ m$; $R = \frac{\pi}{9}$; $r = \sqrt{\min[R^2, (\lambda - \lambda_c)^2 + (\theta - \theta_c)^2]}$, $\lambda_c, \theta_c$ are the center longitude and latitude of the mountain, respectively, we set $\lambda_c = \frac{3\pi}{2}, \theta_c = \frac{\pi}{6}$.

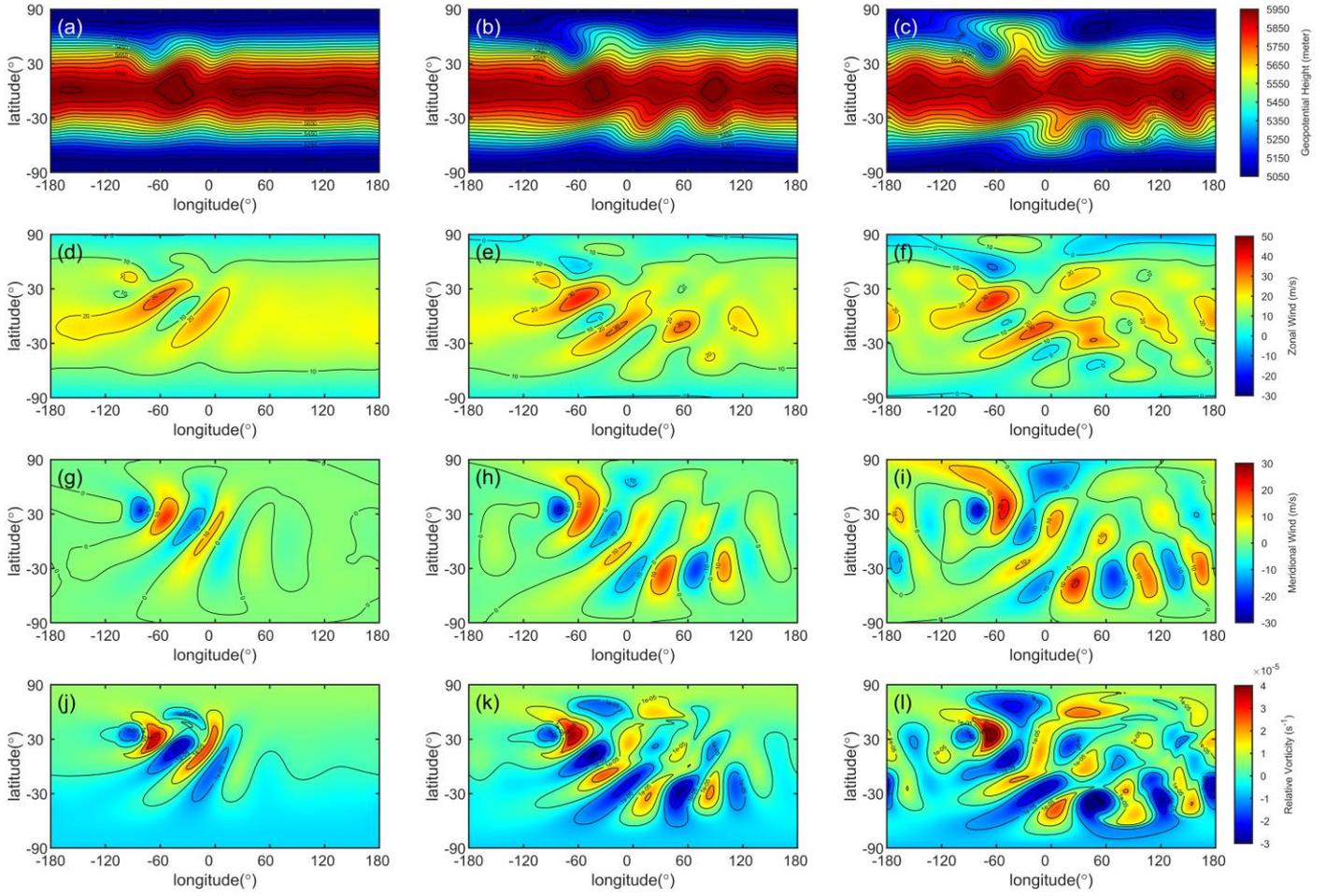

**Figure 5.1** Simulation result of mountain wave on C90 grid. The rows stand for variables: geopotential height, zonal wind, meridional wind and relative vorticity, respectively. The columns represent simulation day 5, 10, 15. Geopotential height contour from 5050 to 5950 $m$ with interval 50 $m$. Zonal wind contour from $-30$ to 50 $m/s$ with interval 10 $m/s$. Meridional wind contour from $-30$ to 30 $m/s$ with interval 10 m/s. Relative vorticity contour from $-3 \times 10^{-5}$ to $4 \times 10^{-5}$ $s^{-1}$ with interval $1 \times 10^{-5}$ $s^{-1}$.

HOPE is able to deal with the bottom topography correctly, as shown in Figure 5.1, all of the simulation result is consistent with prior researches such as [7][27][37] and so on. Furthermore, as discussed in [2], some high order Discontinuous Galerkin (DG) method exhibit non-physical oscillation during simulating the over mountain flow, the additional viscosity operators are necessary to alleviate this issue. However, HOPE does not require any explicit viscosity operator to suppress vorticity oscillations, the vorticity fields are smooth all the time as illustrated in Figure 5.1 (j), (k), (l).

## 5.3 Rossby-Haurwitz Wave with 4 Waves

Rossby-Haurwitz (RH) wave is the 6[th] test case introduced by Williamson et al. (1992)[41], the RH waves are analytic solution of the spherical nonlinear barotropic vorticity equation, the reference solution is the zonal advection of RH wave without pattern changing, the angular phase speed is given by

$$c = \frac{R(R+3)\omega - 2\Omega}{(R+1)(R+2)} \tag{5.9}$$

where $R = 4$ is the zonal wavenumber, $\omega = 7.848 \times 10^{-6} \, s^{-1}$; the earth rotation angular speed $\Omega = 7.292 \times 10^{-5} \, s^{-1}$.

Therefore, we have $c = 29.52\ day$. The initial condition expressed as

$$\phi = \phi_0 + a^2[A(\theta) + B(\theta)\cos R\lambda + C(\theta)\cos 2R\lambda] \tag{5.10}$$

$$u = a\omega\cos\theta + aK\cos^{R-1}\theta\,(R\sin^2\theta - \cos^2\theta)\cos R\lambda \tag{5.11}$$

$$v = -aKR\cos^{R-1}\theta\sin\theta\sin R\lambda \tag{5.12}$$

$$A(\theta) = \frac{\omega}{2}(2\Omega + \omega)\cos^2\theta + \frac{1}{4}K^2\cos^{2R}\theta\,[(R+1)\cos^2\theta + 2R^2 - R - 2 - 2R^2\cos^{-2}\theta] \tag{5.13}$$

$$B(\theta) = \frac{2(\Omega + \omega)K}{(R+1)(R+2)}\cos^R\theta\,[R^2 + 2R + 2 - (R+1)^2\cos^2\theta] \tag{5.14}$$

$$C(\theta) = \frac{1}{4}K^2\cos^{2R}\theta\,[(R+1)\cos^2\theta - R - 2] \tag{5.15}$$

where $\lambda, \theta$ are longitude and latitude, $K = \omega$, $\phi_0 = gh_0$, $h_0 = 8000\ m$, and $a = 6371220\ m$ is the earth radius.

According to the study by Thuburn and Li (2000)[36], the Rossby-Haurwitz (RH) wave with wavenumber 4 is unstable and prone to waveform collapse due to factors such as grid symmetry, initial condition perturbation, and model errors. Similar conclusions have been verified in subsequent research. In tests conducted by Zhou et al. (2020)[47], the TRiSK framework based on the SCVT grid could only sustain the RH wave pattern for 25 days without collapse. In contrast, Li et al. (2020)[18] successfully maintained the RH wave pattern for 89 days using a similar algorithm on a latitude-longitude grid. Ullrich et al. (2010)[37] developed the high-precision MCORE model based on a cubed-sphere grid, which was able to sustain the RH wave for up to 90 days. In the most of our experiments, the ability of HOPE to maintain the Rossby-Haurwitz (RH) wave significantly improved with increased accuracy and grid resolution.

In the 3rd order accuracy simulation, we found that the duration for which the RH wave is maintained increases with higher grid resolution, as exhibit in Figure 5.3. When the grid resolution is low (C45, C90), an obvious dissipation phenomenon can be observed. When the resolution reaches C180, the dissipation is significantly reduced, but the waveform has completely collapsed by day 90. When the resolution reaches C360, the simulation results are further improved, with dissipation further reduced, and the RH wave waveform can still barely be maintained on day 90.

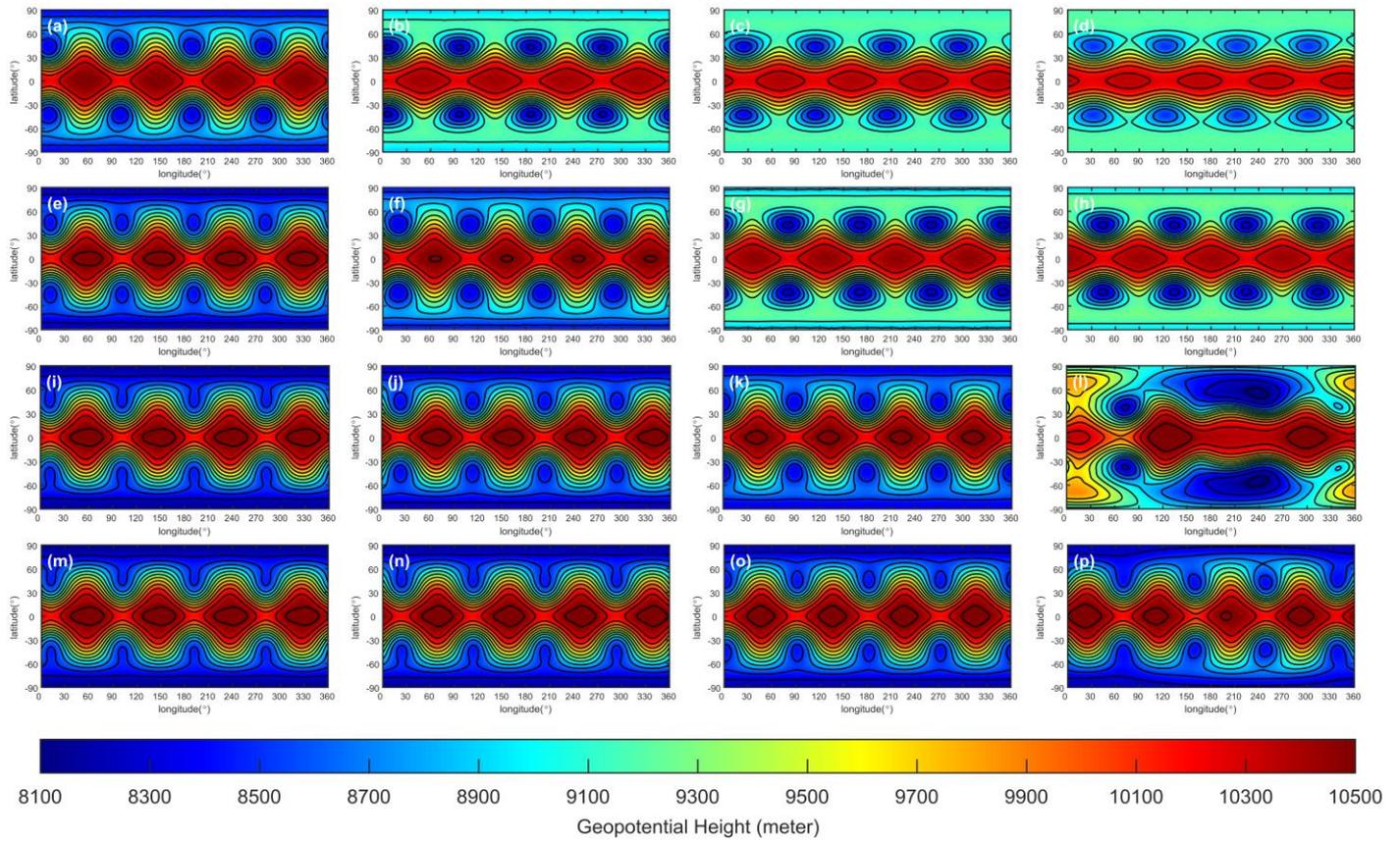

**Figure 5.2** Geopotential height of Rossby-Haurwitz wave simulated by 3$^{rd}$ order spatial reconstruction scheme. The rows represent grid C45, C90, C180 and C360, the columns stand for simulation day 14, 30, 60, 90. Contours from 8100 to 10500 $m$ with interval 200 $m$.

In Figure 5.3, we compare the impact of accuracy on the simulation capability of RH waves by fixing the resolution. By comparing row by row, it can be observed that when the accuracy reaches 5th order or higher, the dissipation is significantly reduced. Both the 5$^{th}$ order and 7$^{th}$ order accuracy simulations show signs of waveform distortion on day 90, and the waveform completely collapses by day 100. However, when using 9$^{th}$ order accuracy for the simulation, the waveform is well maintained even until day 100.

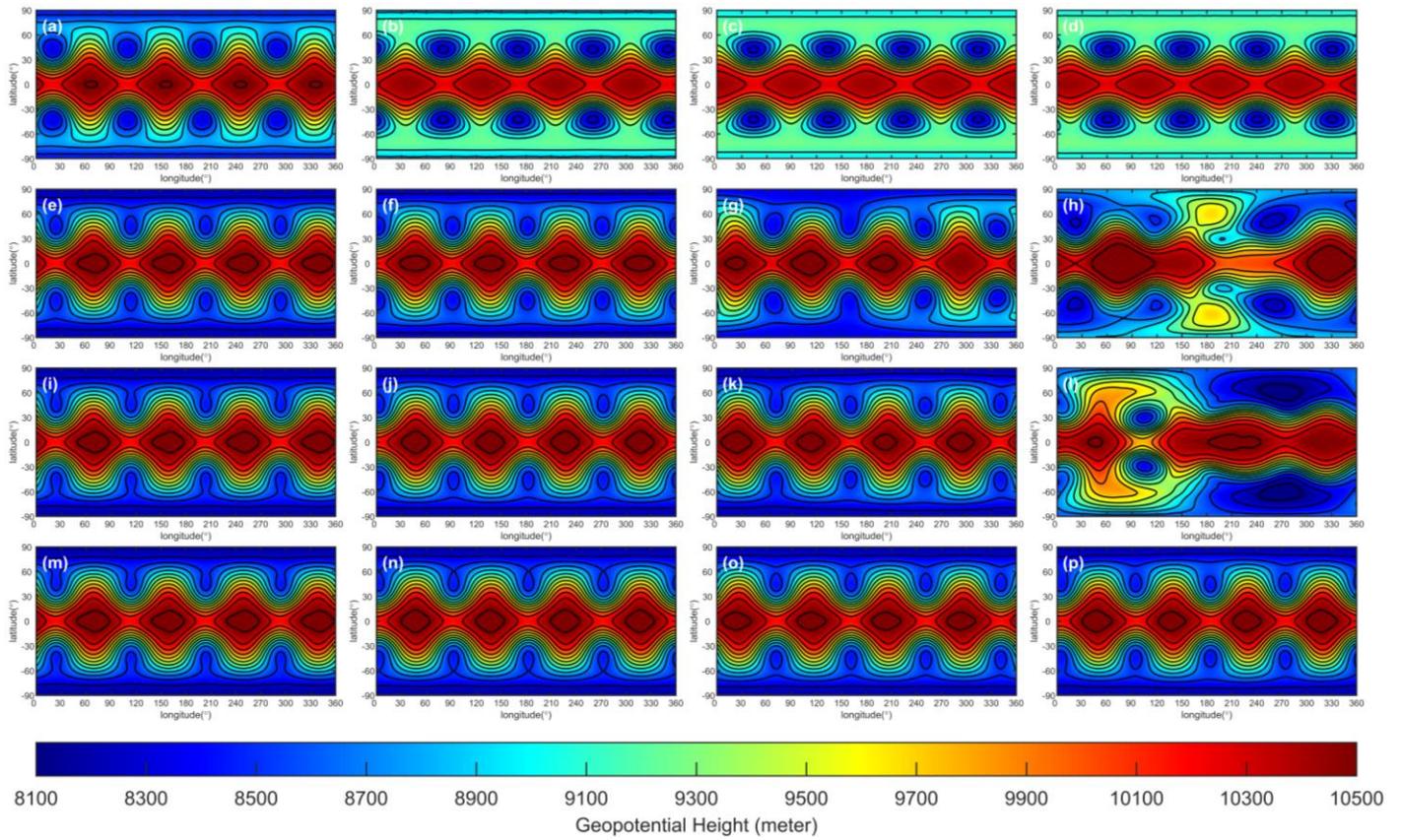

**Figure 5.3** Geopotential height of Rossby-Haurwitz wave on C90 grid, the rows represent the spatial reconstruction scheme with $3^{rd}$, $5^{th}$, $7^{th}$, $9^{th}$ order, the columns stand for simulation day 30, 60, 90 and 100. Contours from 8100 to 10500 $m$ with interval 200 $m$.

Figure 5.4 presents the simulation results on the 80th day for different resolutions and accuracy schemes. The dissipation decreases as the resolution and accuracy improve. At the C45 resolution, both the $3^{rd}$ order and $5^{th}$ order accuracy simulations exhibit significant dissipation. Although the $7^{th}$ order simulation shows a notable improvement in dissipation, the waveform is severely distorted. The $9^{th}$ order accuracy scheme produces the best simulation results. As the resolution increases, the simulation performance also improves significantly. When using the C360 resolution, all accuracy schemes yield good simulation results.

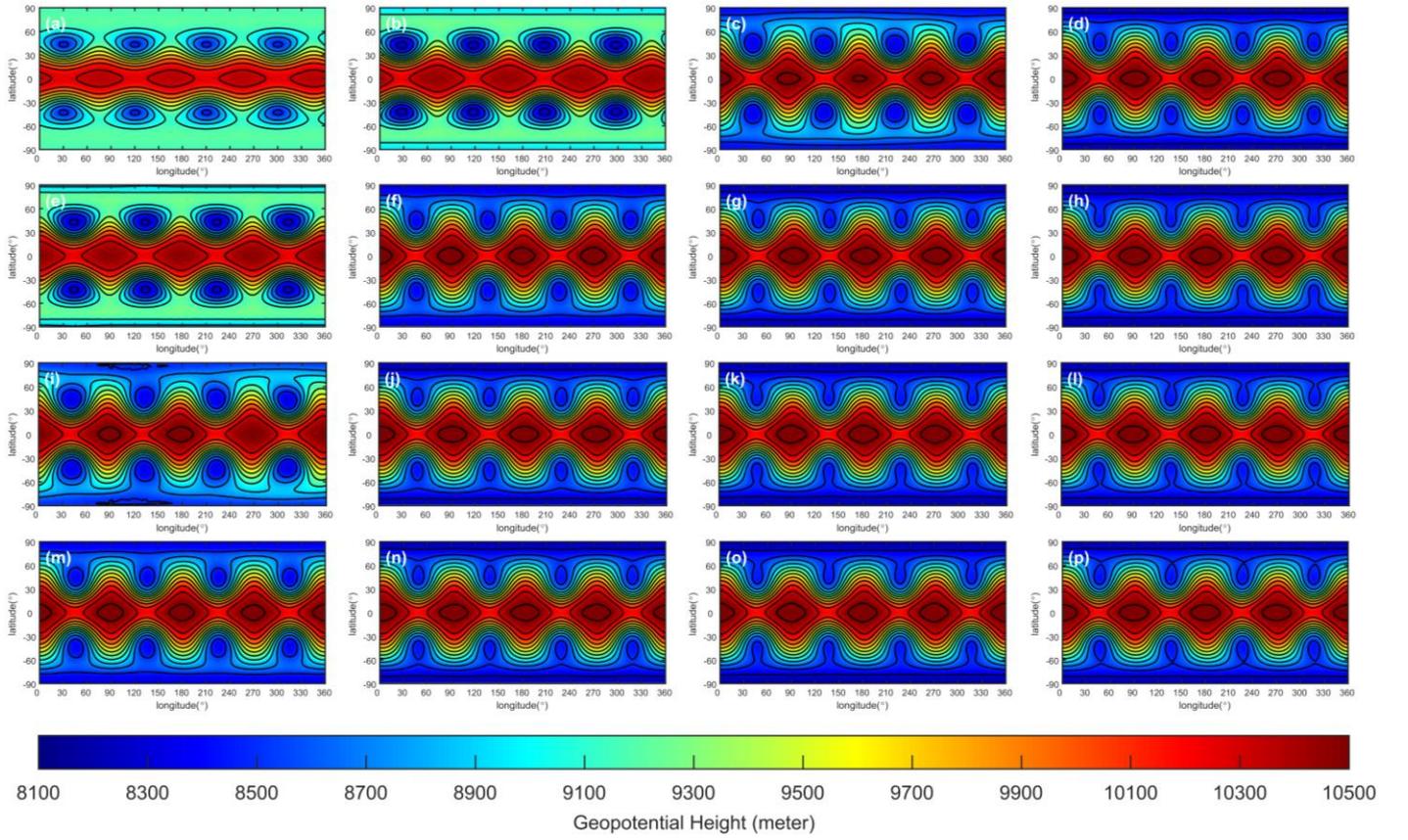

**Figure 5.4** Geopotential height of Rossby-Haurwitz wave at simulation day 80. The rows represent spatial reconstruction with $3^{rd}$, $5^{th}$, $7^{th}$ and $9^{th}$ order. The columns stand for grid C45, C90, C180 and C360. Contours from 8100 to 10500 $m$ with interval 200 $m$.

## 5.4 Perturbed Jet Flow

The perturbed jet flow was introduced by Galewsky et al. (2004)[8], this experiment was desired to test the model ability of simulating the fast and slow motion. the initial field is defined as

$$u(\theta) = \begin{cases} \dfrac{u_{max}}{e_n} e^{\frac{1}{(\theta-\theta_0)(\theta-\theta_1)}}, & \theta \in (\theta_0, \theta_1) \\ 0, & otherwise \end{cases} \quad (5.16)$$

$$\phi(\lambda, \theta) = \phi_0 + \phi'(\lambda, \theta) - \int_{-\frac{\pi}{2}}^{\theta} au(\theta')\left[f + \frac{\tan\theta'}{a}u(\theta')\right]d\theta' \quad (5.17)$$

$$\phi'(\lambda, \theta) = g\hat{h}\cos\theta\, e^{-\left(\frac{\lambda}{\alpha}\right)^2 - \left(\frac{\theta_2-\theta}{\beta}\right)^2}, \lambda \in (-\pi, \pi) \quad (5.18)$$

where $\lambda, \theta$ represents longitude and latitude, $a = 6371220\ m$ is radius of earth, $u_{max} = 80\ m/s$, $\theta_0 = \frac{\pi}{7}, \theta_1 = \frac{5\pi}{14}, \theta_2 = \frac{\pi}{4}$, $e_n = e^{\frac{-4}{(\theta_1-\theta_0)^2}}, \alpha = \frac{1}{3}, \beta = \frac{1}{15}$, and $\hat{h} = 120\ m$.

As mentioned in Chen et al. (2008) [7], the perturbed jet flow experiment poses a particular challenge for the cubed-sphere grid model. Firstly, the jet stream is located at 45°N, which is very close to the boundaries of panel 5 of the cubed-sphere grid, resulting in a large geopotential height gradient in the ghost interpolation region, which leads to larger interpolation error. Furthermore, the location of the geopotential height perturbation $\phi'$ coincides with the boundary between panel 1 and panel 5, which also leads to greater numerical computation errors.

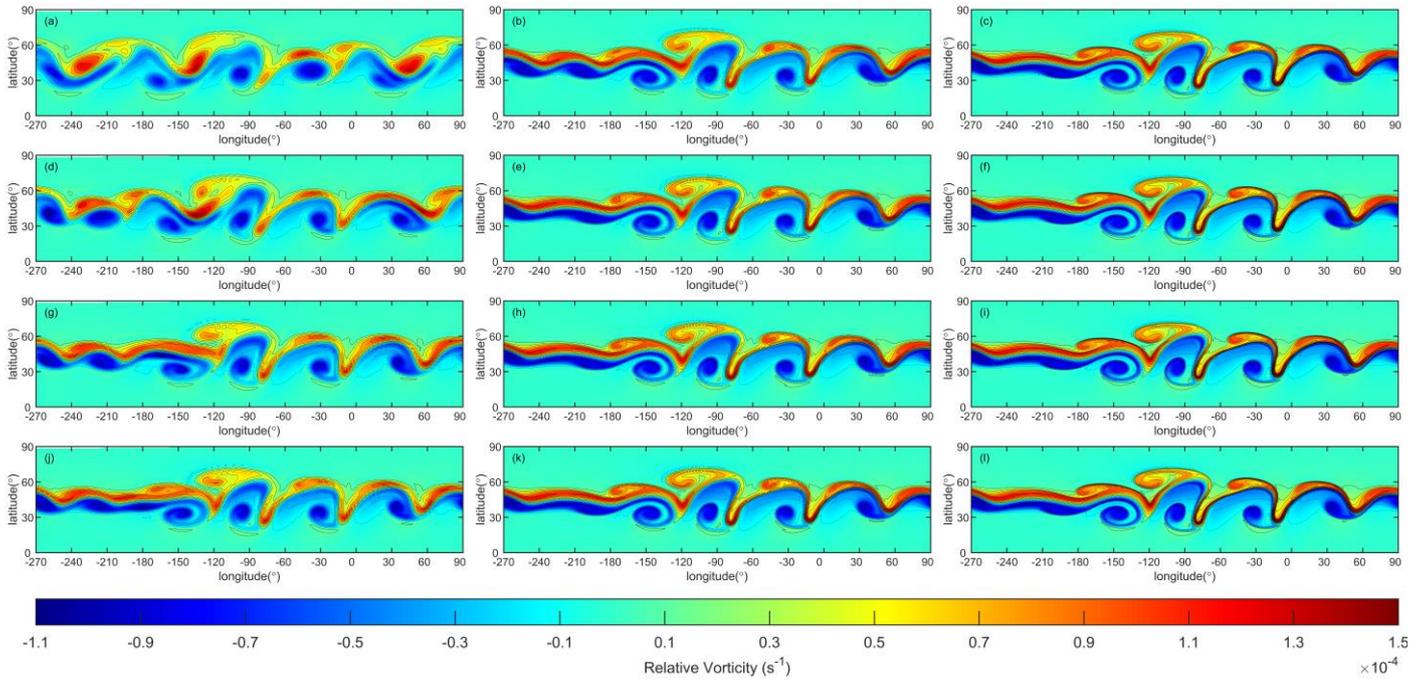

**Figure 5.5** Relative vorticity of perturbed jet flow. (a)~(c) represent the results of 5$^{th}$ order scheme with resolutions C45, C90, C180. (d)~(f) represent the results of 7$^{th}$ order scheme with resolutions C45, C90, C180. (g)~(i) represent the results of 9$^{th}$ order scheme with resolutions C45, C90, C180. (j)~(l) represent the results of 11$^{th}$ order scheme with resolutions C45, C90, C180.

Figure 5.5 displays the HOPE simulation outcomes for varying levels of accuracy and resolutions. The four rows correspond to the 5$^{th}$, 7$^{th}$, 9$^{th}$, and 11$^{th}$ schemes in terms of accuracy. The three columns, meanwhile, represent the resolutions of C45, C90, and C180, respectively. Upon comparing the different columns, it is evident that the perturbed jet flow test case converges as the resolution increases. Figure 5.5 (a), (d), (g), and (j) illustrate that, with an increase in accuracy, the vorticity field patterns become increasingly similar to the high-resolution results shown in the second and third columns of Figure 5.5. Notably, HOPE enhances the simulation results by utilizing both higher accuracy and higher resolution.

## 5.5 Dam-Break Shock Wave

In this section we introduce a dam-break case for testing the capability of HOPE to capture the shock wave and comparing the difference between 1D and 2D WENO schemes. The initial condition is configured as a cylinder with a height of 30000 meters, as shown in Figure 5.6(a). The geopotential height is given by

$$\phi(r(\lambda,\theta)) = \begin{cases} 2\phi_0, & r < r_c \\ \phi_0, & otherwise \end{cases} \quad (5.19)$$

where $r = \sqrt{(\lambda - \lambda_c)^2 + (\theta - \theta_c)^2}, \lambda_c = \pi, \theta_c = 0, r_c = \frac{\pi}{9}, \phi_0 = gh_0, h_0 = 30000\ m$, and the earth rotation angular speed $\Omega = 0$.

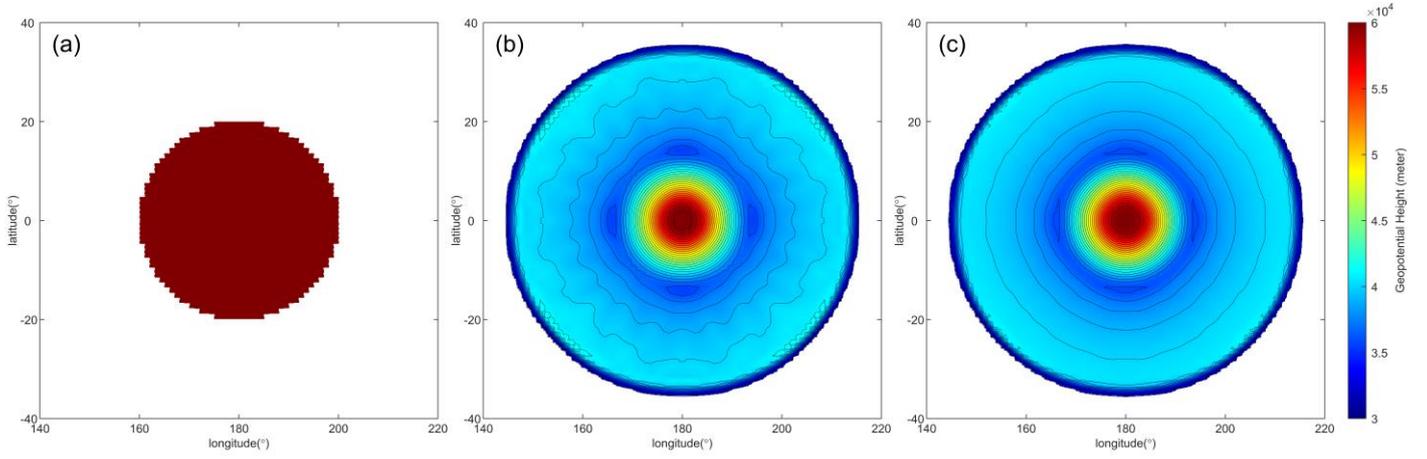

**Figure 5.6** Geopotential height of dam-break test case on C90 grid at 2nd hour. (a) Initial condition, (b) WENO 1D, (c) WENO 2D. The horizontal resolution for both schemes is C90. Shaded and contour from $3.2 \times 10^4$ to $6 \times 10^4$ meters, with contour interval $10^3$ meters.

In this experiment, we adopt $5^{th}$ order accuracy for both 1D and 2D schemes, the WENO-Z [4] is adopted as WENO 1D scheme, and WENO 2D scheme is consist with section 3.2. Due to the initial condition being a cylinder, the resulting shock wave should maintain a circular feature. In the simulation results of WENO 1D, numerous radial textures appear, Figure 5.6(b). The simulation results using the WENO 2D scheme exhibit a smoother circular shape, Figure 5.6(c). This outcome arises because the 1D reconstruction scheme suffers from dimension split error, whereas the fitting function in the 2D reconstruction scheme incorporates cross terms, significantly improving the handling of anomalous anisotropic characteristics.

# 6. Conclusions

In this article, we present HOPE, an innovative high-order finite volume model that boasts the capability of achieving arbitrary odd-order convergence. When the flow fields are sufficiently smooth, the high-order accuracy characteristics of HOPE enable rapid convergence of simulation errors. In simulation experiments of Rossby-Haurwitz waves, HOPE's ability to maintain the waveform improves as accuracy order and spatial resolution increase. Similarly, for the perturbed jet flow case, HOPE successfully captures both fast and slow motion features in the flow field, and the simulation results are significantly improved as the accuracy order and spatial resolution increase. In simulations of Mountain waves, HOPE accurately processes gravity waves induced by bottom topography. In the dam break case, where a cylindrical shock front is present, the WENO algorithm with a two-dimensional reconstruction scheme outperforms the dimension split scheme in maintaining the circular feature.

Moreover, the algorithm of HOPE can effectively leverage the performance of GPUs. Spatial reconstructions are implemented using convolution operators, and integration operations are equivalent to matrix-vector multiplications, both of which are widely utilized in the field of machine learning. Additionally, HOPE has been developed using PyTorch, thereby inherently benefiting from its automatic differentiation capability. Notably, HOPE has been developed using PyTorch, thereby inheriting its automatic differentiation capability. This seamless integration allows HOPE to be effortlessly combined with any neural network (NN) system, paving the way for the construction of a hybrid prediction model that merges a high-order, high-performance numerical dynamic core with an NN-based physical parameterization package.

In our ongoing research, we have implemented a similar algorithm to develop a two-dimensional baroclinic model. Looking ahead, we plan to leverage the HOPE algorithm to create a global, fully compressible baroclinic model, highlighting the innovative strength and advantage of this algorithm in modeling baroclinic systems.

# 7. Appendix

In this appendix, we introduce a novel boundary ghost cell interpolation scheme for cubed sphere, which is able to support HOPE to reach the accuracy over 11$^{th}$ order or even higher.

There are two types of cells, in-domain and out-domain (also named ghost cell, as show in Figure 3.6(b)), we define the set of in-domain cell values $\boldsymbol{q}_{d\times1} = (q_1, q_2, \ldots, q_d)^T$, the set of out-domain cell values $\boldsymbol{g}_{h\times1} = (g_1, g_2, \ldots, g_d)^T$, and the set of Gaussian quadrature point values (green points in Figure 3.2) in out-domain cells is define as $\boldsymbol{v}_{p\times1} = (v_1, v_2, \ldots, v_p)$. To identify the shape of the arrays, we denote the array shape using subscripts (this convention will be followed throughout the subsequent text). The purpose of ghost cell interpolation is using the known cell value $\boldsymbol{q}$ to interpolate the unknown $\boldsymbol{g}$.

Define a new set includes the values of domain cell values and ghost cell values

$$\tilde{\boldsymbol{q}}_{(d+h)\times1} = \boldsymbol{q} \cup \boldsymbol{g} = (q_1, q_2, \ldots, q_d, g_1, g_2, \ldots, g_h)^T \tag{7.1}$$

Similar to the describe in section 3.1, we can use a TPP to reconstruct the ghost quadrature points

$$\boldsymbol{v}_{p\times1} = A_{p\times(d+h)} \tilde{\boldsymbol{q}}_{(d+h)\times1} \tag{7.2}$$

where $A_{p\times(d+h)}$ is the interpolation matrix that can be obtain by the similar method to (3.11). The ghost cell values are calculated by Gaussian quadrature

$$\boldsymbol{g}_{h\times1} = B_{h\times p} \boldsymbol{v}_{p\times1} \tag{7.3}$$

where $B_{h\times p}$ is the Gaussian quadrature matrix.

$\tilde{\boldsymbol{q}}_{(d+h)\times1}$ can be decomposed as the linear combination of $\boldsymbol{q}_{d\times1}$ and $\boldsymbol{v}_{p\times1}$

$$\tilde{\boldsymbol{q}}_{(d+h)\times1} = \begin{pmatrix} I_{d\times d} & 0 \\ 0 & B_{h\times p} \end{pmatrix} \begin{pmatrix} \boldsymbol{q}_{d\times1} \\ \boldsymbol{v}_{p\times1} \end{pmatrix} = \tilde{B}_{(d+h)\times(d+p)} \overline{\boldsymbol{q}}_{(d+p)\times1} \tag{7.4}$$

where $I_{d\times d}$ is an identity matrix, and

$$\tilde{B}_{(d+h)\times(d+p)} = \begin{pmatrix} I_{d\times d} & 0 \\ 0 & B_{h\times p} \end{pmatrix} \tag{7.5}$$

$$\overline{\boldsymbol{q}}_{(d+p)\times1} = \begin{pmatrix} \boldsymbol{q}_{d\times1} \\ \boldsymbol{v}_{p\times1} \end{pmatrix} \tag{7.6}$$

Substitute Eq.(3.12) into Eq.(3.8), we have

$$\boldsymbol{v}_{p\times1} = A_{p\times(d+h)} \tilde{B}_{(d+h)\times(d+p)} \overline{\boldsymbol{q}}_{(d+p)\times1} = \tilde{A}_{p\times(d+p)} \overline{\boldsymbol{q}}_{(d+p)\times1} = \tilde{A}_{p\times(d+p)} \begin{pmatrix} \boldsymbol{q}_{d\times1} \\ \boldsymbol{v}_{p\times1} \end{pmatrix} \tag{7.7}$$

We found that matrix $\tilde{A}_{p\times(d+p)}$ can be decomposed into two parts

$$\tilde{A}_{p\times(d+p)} = \begin{pmatrix} \overline{A}_{p\times d} & C_{p\times p} \end{pmatrix} \tag{7.8}$$

Such that
$$\boldsymbol{v}_{p\times 1} = \bar{A}_{p\times d}\boldsymbol{q}_{d\times 1} + C_{p\times p}\boldsymbol{v}_{p\times 1} \tag{7.9}$$

Therefore
$$(I_{p\times p} - C_{p\times p})\boldsymbol{v}_{p\times 1} = \bar{A}_{p\times d}\boldsymbol{q}_{d\times 1} \tag{7.10}$$

We set $D_{p\times p} = I_{p\times p} - C_{p\times p}$, then $\boldsymbol{v}_{p\times 1}$ can be determined by
$$\boldsymbol{v}_{p\times 1} = D_{p\times p}^{-1}\bar{A}_{p\times d}\boldsymbol{q}_{d\times 1} \tag{7.11}$$

Substitute Eq.(7.11) into Eq.(7.3), we establish the relationship between ghost cell values and in-domain cell values
$$\boldsymbol{g}_{h\times 1} = B_{h\times p}\boldsymbol{v}_{p\times 1} = B_{h\times p}D_{p\times p}^{-1}\bar{A}_{p\times d}\boldsymbol{q}_{d\times 1} = G_{h\times d}\boldsymbol{q}_{d\times 1} \tag{7.12}$$

where $G_{h\times d} = B_{h\times p}D_{p\times p}^{-1}\bar{A}_{p\times d}$. It's clear that Eq.(7.12) is linear, and only rely on the mesh and Gaussian quadrature scheme. Therefore, we need to compute the projection matrix $G_{h\times d}$ only once for a given mesh and accuracy, this matrix can be computed by a preprocessing system and save it to the hard disk.